\documentclass[aps,prx,two column,showpacs,amsmath,amssymb,superscriptaddress]{revtex4-2}
\usepackage{graphicx}
\usepackage{dcolumn}
\usepackage{bm}
\usepackage{xcolor}
\usepackage{amsmath}
\usepackage{pgfplots}
\usepackage{algorithm}
\usepackage{amssymb} 
\usepackage{caption}
\usepackage{float}  
\usepackage{booktabs}
\usepackage{comment}

\begin{document}

\title{Predicting liquid properties and behaviour via droplet pinch-off and machine
learning}

\author{Jingtao Wang}
\thanks{These authors contributed equally to this work.}
\affiliation{Department of Mechanical Engineering, University College London, Torrington Place, London, WC1E 7JE, United Kingdom}

\author{Qiwei Chen}
\thanks{These authors contributed equally to this work.}
\affiliation{Mechanical Science \& Engineering, University of Illinois Urbana-Champaign, Urbana, IL 61801, United States of America}

\author{C.~Ricardo Constante-Amores}
\affiliation{Mechanical Science \& Engineering, University of Illinois Urbana-Champaign, Urbana, IL 61801, United States of America}

\author{Denise Gorse}
\affiliation{Department of Computer Science, University College London, London, WC1E 6EA, United Kingdom}

\author{Alfonso Arturo Castrej\'on-Pita}
\email{alfonso.castrejon-pita@wadham.ox.ac.uk}
\affiliation{Department of Engineering Science, University of Oxford, Parks Road, Oxford OX1 3PJ, United Kingdom}

\author{Jos\'e Rafael Castrej\'on-Pita}
\email{r.pita@ucl.ac.uk}  
\affiliation{Department of Mechanical Engineering, University College London, Torrington Place, London, WC1E 7JE, United Kingdom}

\date{\today}

\begin{abstract}

Here, we demonstrate that the time-evolving interface during droplet formation—and, in particular, the morphology near pinch-off—encodes sufficient physical information for machine-learning (ML) models to accurately infer key fluid properties, including viscosity and surface tension. A dataset was constructed from high-speed imaging of droplet formation in both dripping and \textcolor{black}{and jetting} regimes. The frame closest to break-up was selected and the contour of the droplet extracted and recorded, along with the corresponding flow conditions and fluid properties. Experiments were conducted using Newtonian fluids, such as silicon oils, aqueous solutions of ethanol and glycerin, and methanol, under controlled conditions, spanning 0.001 \(< \) Re \(< \) 200 and 0.01 \(< \) Oh \(< \) 20, providing 840 samples. Supervised regression models were trained to predict fluid properties from these inputs. The models accurately infer surface tension and viscosity across a wide range of fluids using the single high-speed snapshot. Moreover, apart from predicting fluid properties from a snapshot, the proposed framework can carry out the inverse problem and predict droplet shape at break-up based on fluid properties and flow conditions. Unsupervised clustering of the learned representations reveals distinct regions in the Re–Oh and Bo–Oh parameter spaces, indicating that the latent space captures meaningful physical structure and provides insight into droplet dynamics. These results establish a data-driven alternative to conventional measurement techniques, reducing experimental complexity and evaluation time while enabling integration into automated systems.

\end{abstract}

\maketitle

\section{Introduction}
The quantification of liquid properties, such as surface tension, viscosity, and density, is an essential quality-control measure in fluid dispensing processes such as inkjet printing\cite{dearden2005low, stringer2009limits} and spray coating \cite{pan2018coatings, carey2018spray}. However, some conventional methods for determining fluid properties rely on specialized and complex instrumentation, which often limits their use within small enterprises and complicates automation. Among various existing techniques, the \emph{pendant drop} method has become a widely adopted approach to determine the surface tension of a liquid \cite{stauffer1965measurement}. In this method, a droplet is suspended from the tip of a capillary, or needle, with the droplet contour recorded by optical imaging. The shape of the pendant droplet, and thus its contour, is controlled by a balance between gravity and interfacial forces \cite{saad2011design}. Once the droplet boundary is recorded, numerical solutions of the Young–Laplace equation, which accounts for the pressure difference across the liquid interface, the surface curvature, and surface tension \cite{bashforth1883attempt}, are fitted to determine the surface tension $\sigma$ \cite{hansen1991surface}. This method is valued for its simplicity, small liquid volume requirements, and compatibility with conventional optical imaging systems, making it suitable for both research and industrial environments \cite{berry2015measurement}. \textcolor{black}{In practice, millimetre-scale droplets are commonly employed in this method \cite{lin1995examination,lin1996systematic,morita2002influence}}. Other experimental techniques used to characterise surface tension include the Wilhelmy plate, the Du No\"uy ring, and the widely used bubble pressure tensiometry.  In the Wilhelmy plate method the surface tension of a liquid is determined by measuring the force exerted by the liquid on a thin fully-wetted plate in contact with its surface \cite{harkins1930method}.  Alternatively, and in a similar fashion, the Du No\"uy ring method works by lifting a thin metallic ring from a liquid surface while recording the force required to detach it \cite{harkins1930method}. These two methods are simple, but their accuracy depends on the skills of the operator and on the use of materials that ensure the complete wetting of the plate or ring. On the other hand, bubble pressure tensiometers are reliable systems that obtain surface tension by measuring the pressure required to produce a bubble at the tip of a capillary submerged in the tested liquid \cite{simon1851recherches}. In this method, the bubble production rate is adjustable, making the technique  particularly effective for studying time-dependent interfacial phenomena, such as surfactant diffusion and adsorption \cite{dixit2012application}. However, bubble pressure tensiometers are usually expensive and less suitable for highly viscous fluids \cite{fainerman2004maximum}.

Viscosity is another key property that controls various aspects of fluid mechanics on the capillary scale, including jetting\cite{weber1931zerfall}, spray dynamics\cite{gaillard2022determines}, and liquid break-up \cite{castrejon2015plethora}. In industrial environments, viscosity is often measured by vibrational viscometers and rheometers. Vibrational methods work by quantifying the viscous damping of shear acoustic waves through the liquid. These viscometers are fast, easy to operate, and provide reliable results for liquids in which viscosity remains constant regardless of shear; these liquids are commonly referred to as `Newtonian fluids'. A disadvantage of vibrometers is that they require large volumes of liquid, often in the 200 ml range. Liquids with shear-dependent viscosity, such as shear-thinning or thickening, and viscoelastics, require capillary and/or rotational rheometers \cite{schramm1994practical}. These rheometers determine viscosity by applying a controlled shear stress, measuring the resulting flow response of the fluid. Unfortunately, rheometers are often both expensive and unsuitable for the fast-shear rates found in sprays and inkjet printing \cite{koehler2005new,castrejon2021formulation}. 

Whilst effective, the traditional techniques described above form a segmented toolkit that lacks integration and are often incompatible with automation. In this work, we present an integrated, automation-friendly methodology, utilizing both supervised and unsupervised machine learning techniques, that can determine the surface tension, or the viscosity, or both, of Newtonian fluids from a single snapshot of a dripping drop. In our methods, the flow rate, the nozzle size, the density, and, if available, the viscosity or the surface tension are used as additional input. 

In more detail, high-resolution and high-speed imaging experiments were conducted in this work to capture the last moments before the break-up of dripping droplets. From these images, droplet contours were extracted and used as key input features to supervised learning models. Various models are then trained to predict physical parameters, such as viscosity and surface tension, directly from the image data. In addition, models are trained to predict the break-up shape of liquids given their material properties and flow rate. Unsupervised clustering algorithms are additionally employed to discover latent structures within the dataset, enabling the identification of distinct droplet behaviours, or quality modes, without requiring predefined labels. ML methods require sufficient and representative datasets, along with well-defined input features and target outputs; here, the training is carried out directly from  experimental data, taken under a large variety of conditions, ensuring that the learned models capture the intrinsic complexity of wide laboratory conditions. This dual (supervised plus unsupervised) approach not only offers a fast and automated alternative to conventional rheometry and tensiometry but also reveals interpretable morphological patterns linked to fluid properties and experimental conditions. Ultimately, our integration of physics-informed  imaging (based on our hypothesis) and machine learning offers a scalable and generalizable pathway for data-driven fluid diagnostics. \textcolor{black}{Importantly, as it will be discussed later, our approach relies on the finite-time morphology observed during experimentally accessible pinch-off dynamics, rather than on the ideal asymptotic similarity solutions that arise close to the singularity.}

\section{Background} \label{sec:Background}

\subsection{Liquid break-up} \label{sec:LiquidBreakup}
For over a century, the break-up of liquid jets and filaments has been at the forefront of fluid mechanics, given its relevance to a myriad of processes ranging from lava ejection and fragmentation to inkjet printing \cite{castrejon2012breakup, basaran2013nonstandard}. This phenomenon, while seemingly simple, involves complex dynamics where a liquid thins, or necks, rapidly until pinch-off, evolving through dramatic interfacial changes, all within a few microseconds \cite{basaran2002small}. Indeed, the point of break-up is a finite-time singularity and its description requires both scaling arguments and high-resolution experiments and modelling to resolve features that arise in its vicinity. A comprehensive theoretical framework has classified the routes to break-up into various canonical regimes depending on the relative importance of inertia, capillarity, and viscosity \cite{basaran2002small}. In the inviscid regime, where inertia balances surface tension, the thinning of a liquid neck follows the power law $h_{\text{min}} \sim (t_b - t)^{2/3}$, where $h_{\text{min}}$ is the minimum neck radius, $t_b$ is the break-up time and $t$ is time. Interestingly, liquid break-up under this regime shows self-similarity where, regardless of the flow conditions, the liquid interface near pinch-off converges toward a unique shape, offering a powerful criterion for identifying inviscid liquids\cite{day1998self, castrejon2012self}. This regime is often observed in low–viscosity fluids such as water or ethanol, when surrounded by a negligible ambient environment\cite{keller1983surface,chen2002computational}. \textcolor{black}{Previous studies have attempted to infer material properties from pinch-off dynamics by fitting the similarity law  $R_{\min} \sim A (t_b - t)^{2/3}$, using a theoretical prefactor $A\approx 0.717$. However, further experiments have demonstrated that this approach can lead to significant errors in the inferred surface tension as experimental observations often occur outside the self similarity universal regime  \cite{hauner2017dynamic}. Subsequent work showed that the prefactor exhibits a slow temporal drift and depends on system parameters in this pre-asymptotic regime \cite{deblais2018viscous}. We also note that convergence to the self-similar regime near the capillary singularity is slow and involves multiple transient regimes \cite{deblais2018viscous,Corpart_Herrada_Deblais_Bonn_2025}, and rarely achieved in experiments. As noted by \cite{paulsen2012inexorable}, different regions may exhibit distinct force balances, so the route to break-up is parameter-dependent and the morphology reflects this history.  Our approach takes advantage of the time history of the dynamics, which remains imprinted in the interface topology up to pinch-off. In summary, asymptotic similarity eliminates independent parameter information in the strict infinite-time limit, and only around the pinch off region. Real finite-time experiments retain sufficient discriminatory structure to allow the separate inference of $\sigma$ and $\mu$. Our approach makes use on the interface morphology rather than  predicted prefactors.}

In most other liquids, both viscosity and inertia play a dominant role. In fact, when viscous forces dominate, the thinning is characterized by a linear scaling, i.e. $h_{\min} =0.0709 \left( \frac{\mathrm{\sigma}}{\mathrm{\mu}} \right) (t_b - t)$, where $\sigma$ is surface tension and $\mu$ is the viscosity\cite{papageorgiou1995breakup}. In contrast, in cases where all three forces, i.e. capillary, viscous, and inertial, play a role in the dynamics, the viscous–inertial regime  leads to the universal scaling $h_{\min} = 0.0304\left( \frac{\mathrm{\sigma}}{\mathrm{\mu}} \right) (t_b - t)$\cite{eggers1993universal}.  A further fourth regime exists where the surrounding media viscosity, often another liquid, dictates the thinning and pinch-off dynamics  \cite{li2016capillary}. It is worth noting that, on the route to break-up, the neck dynamics transition between these regimes based on the local importance of inertia, viscosity and surface tension effects \cite{castrejon2015plethora}. The thinning regimes, and their transitions, are parametrized by two dimensionless parameters, the Reynolds ($Re=\frac{4 \rho Q}{ \pi \mu D}$), and the Ohnesorge ($Oh=\frac{\mu}{\sqrt{\rho \sigma D}}$) numbers, where $Q$ is the volumetric flow rate, $\rho$ is the liquid density, and $D$ is a characteristic length (often the nozzle size or the  neck diameter). In fact, the transitional behavior has been found to consist of multiple intermediate transient regimes, which are highly sensitive to the liquid and flow properties. In other words, in their route to break-up, two liquids of similar characteristics can have a very different thinning transitional behavior \cite{castrejon2015plethora}. Thus, the profile of a liquid neck near, or at, break-up is a repository of fluid dynamics information, reflecting the transitional competition of fluid properties and the flow at the neck. Complementary works have demonstrated that the  break-up of liquid filaments, and the transition from dripping to jetting, are also ruled by the Reynolds and Ohnesorge numbers and by the filament length to diameter aspect ratio\cite{ambravaneswaran2004dripping,notz2004dynamics}.  
\textcolor{black}{These findings suggest that the route to break-up is unique for each combination of flow characteristics and liquid properties, and thus governed by their complex interplay. Motivated by this, we have assembled a large database of snapshots of droplets undergoing breakup (dripping) train ML models for liquid-property (and interface) inference.}

\subsection{Machine learning}

Machine learning (ML) is a branch of computer modeling that learns patterns from large amounts of data and makes predictions, or decisions, without prior knowledge beyond what it has learned from this data. In recent years, ML has been applied across various physics topics, including fluid dynamics\cite{ALVERINGH2023114762,chen2025dynamicsdatadrivenlowdimensionalmodel,constanteamores2025dynamicsdatadrivenlowdimensionalmodel}, droplet physics \cite{gaikwad2022process,doi:10.34133/research.0197,Kratz_2020,li2025machine, kim2022design}, soft matter \cite{clegg2021characterising,orlova2023machine}, and materials science \cite{MOBARAK2023100523}. These studies highlight the potential of ML as a tool to complement traditional experimental and theoretical approaches in science. Specifically, supervised learning trains models on labelled datasets (data that has been manually classified or characterised), mapping inputs to known outputs, which is particularly effective for prediction tasks. In contrast, unsupervised learning operates on unlabelled data to identify hidden patterns or underlying structures, i.e. clusters. Both strategies offer valuable means of extracting knowledge from complex experimental observations. In addition, recent advances in computer vision\cite{masci2011stacked} provide powerful tools for analysing image-based experimental data \cite{khor2019using}, which are highly relevant for the study of droplet dynamics and pinch-off processes where high-speed imaging is often used in conjunction with image analysis to extract information from images\cite{castrejon2015plethora}.

\subsection{ML Applications to Droplet Dynamics} \label{sec:RelatedWork}

ML methods have been applied to various systems involving droplet formation, examples of these include direct inkjet writing\cite{zhang2021reviews}, droplet impact, spreading and bouncing on nanostructure substrates\cite{au2023predicting}, spreading on various materials \cite{tembely2022machine}, and droplet splashing \cite{pierzyna2021data}. Considering first the forward-prediction paradigm, in which ML models predict droplet behaviour from known material and operating conditions, Kim et al.\cite{kim2023predicting} trained supervised regressors to predict the jetting behaviour of viscoelastic inks from rheological and drive parameters, with later work \cite{kim2025reinforcement} using reinforcement learning to optimise actuation waveforms, dynamically improving jet stability without an explicit physical model. Additionally, Chen et al.\cite{doi:10.34133/research.0197} employed supervised learning to predict the optimum printable biomaterial formulations for direct ink writing, and Li et al.\cite{li2025machine} developed a multi-parameter droplet optimisation model including flow and material parameters. Turning to image-based studies, Kratz and Kierfeld \cite{Kratz_2020} introduced an image-based ML approach for pendant-drop tensiometry, demonstrating that the contour of a static droplet can be used to infer surface tension with the method restricted to steady pendant configurations. Pierzyna et al. \cite{pierzyna2021data} developed a data-driven model for the splashing thresholds and Au-Yeung et al. \cite{au2023predicting} developed a method to predict the maximum spreading and impact morphology of droplets on nanostructured surfaces. 

In summary, previous works have predominantly used ML to control, or classify, droplet behaviour given known physical inputs. In contrast, the present work uses supervised learning to tackle the inverse problem of inferring fluid properties directly from droplet morphology. In addition, we use unsupervised learning to gain insight into the underlying physics of this process.

\section{Experimental Methodology} \label{sec:ExperimentalMethodology}

In our experiments, dripping droplets were produced by pushing liquid, at a controlled flow rate, through long cylindrical nozzles. The flow rate was controlled by a KDS Gemini 88 Plus Dual Rate syringe pump, through Hamilton glass syringes (models 1750, 1001, and 1005; with nominal volumes of 0.5, 1.0 and 5.0 ml, respectively). Polyurethane tubing, of 6.0 mm diameter, connected the syringes to the nozzles by luer adaptors. The following flow rates were used: 100, 250, 500, 750, 1,000, 1,250, 1,500, 1,750, and 2,000 $\mu$l/min. The actual maximum flow rate per liquid was limited by the fluid viscosity, e.g. the maximum flow rate achieved for pure glycerine was $Q=1,000$ $\mu$l/min. Six different nozzles were used in our experiments, with internal diameters of 0.10, 0.25, 0.83, 1.37, 1.79, and 2.38 mm. Prior to data acquisition, the liquid delivery system was carefully primed to minimise air entrapment. During experiments, the pump imposed a constant volumetric flow rate, producing a pendant drop that grew under gravity, underwent necking, and pinched off. The complete process, from necking to detachment, was recorded for subsequent frame-by-frame  (visual) analysis to identify the frame nearest break-up.

\begin{figure*}
    \centering
   \includegraphics[width=0.7\textwidth]{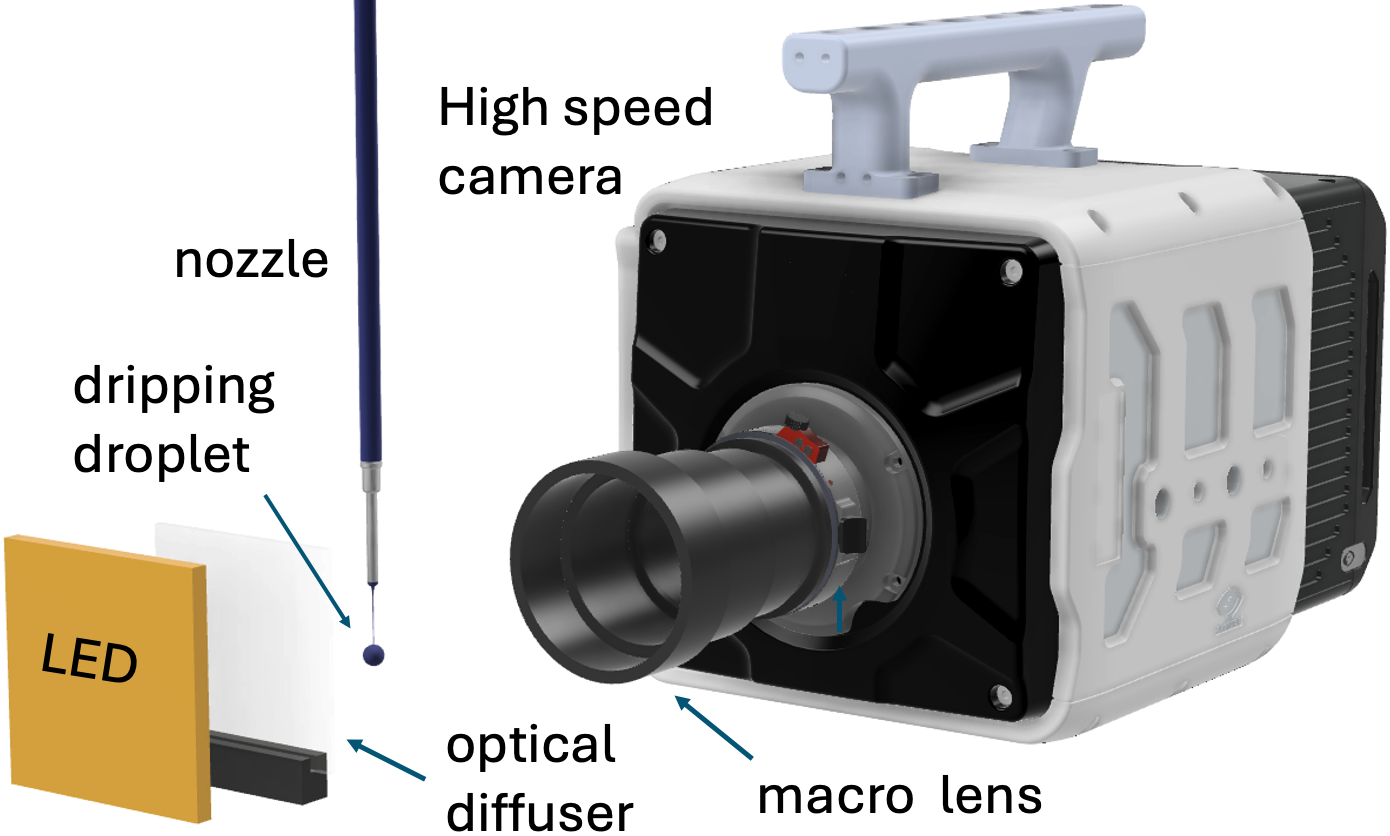}
    \caption{Experimental setup used to capture the necking and break-up dynamics during droplet dripping. The system is a simple shadowgraph, where  visualisation was carried out through a macro lens and a diffused light source.} 
    \label{fig:Experimental setup}  
\end{figure*}

\begin{figure*}
\includegraphics[width=0.4\linewidth]{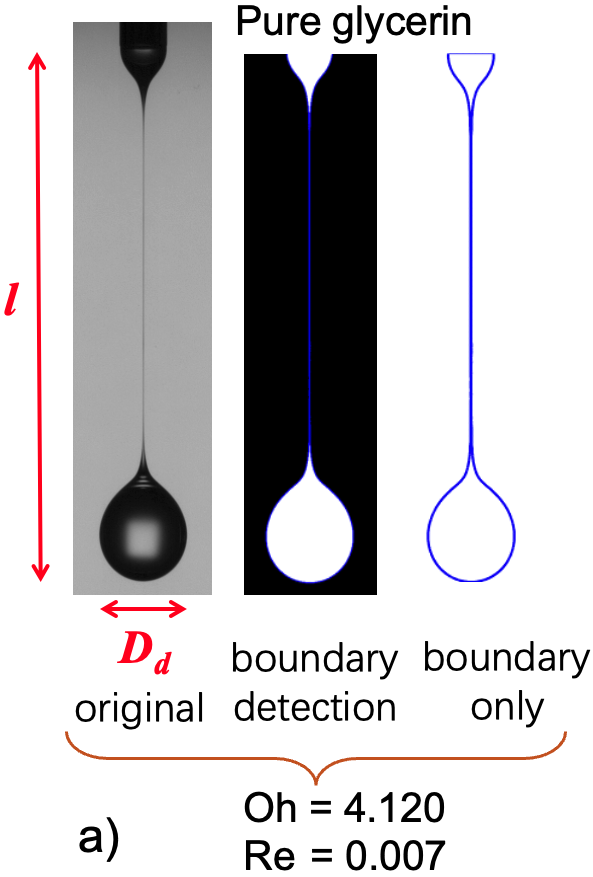}
\includegraphics[width=0.58\linewidth]{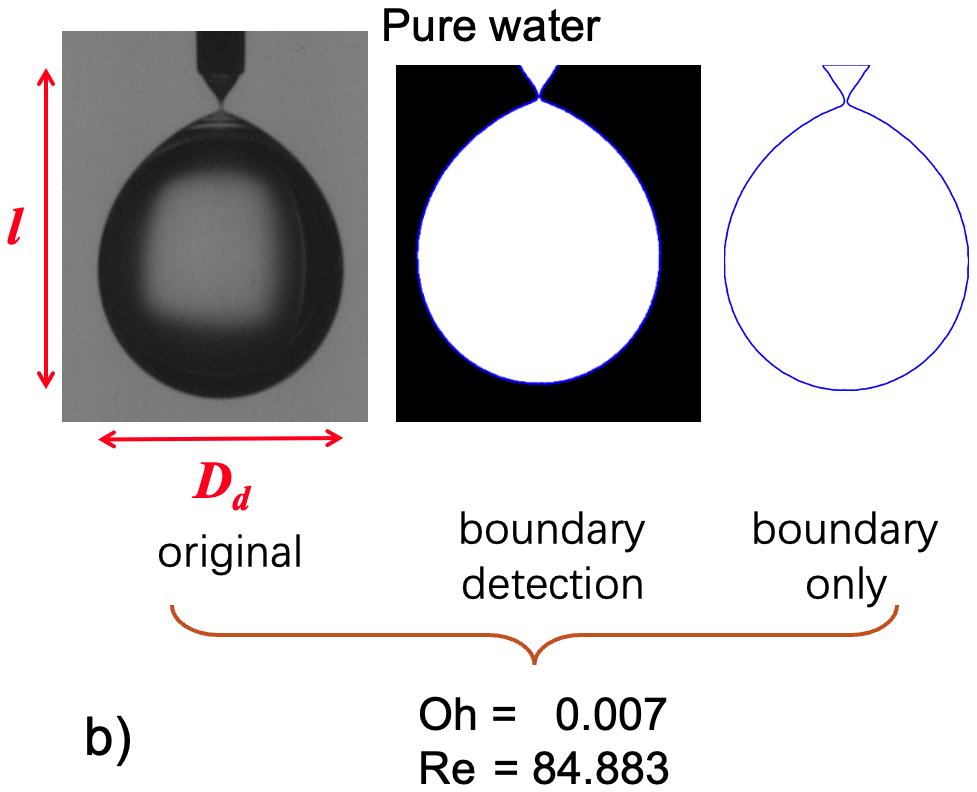}
\caption{Examples of the image-processing pipeline. The left panel shows the droplet at break-up, 
the middle panel shows the binary segmentation, and the right panel shows the extracted interface silhouette.}\label{fig:liquids_and_processing}
\end{figure*}

Drop formation and pinch-off were recorded using the back-lit, side-view, high-speed imaging system illustrated in Fig.~\ref{fig:Experimental setup}. The setup consists of a high-speed camera (Phantom, TMX-5010), an off-the-shelf macro lens (Tamron macro), a vertical nozzle for gravity-driven dripping, a 500 Watt LED light, and an optical diffuser. The camera–lens axis was aligned orthogonally to the nozzle to obtain the side view. The diffuser is used to produce a uniform background and high-contrast images; some examples are seen in Figures \ref{fig:liquids_and_processing}. Images were acquired at a spatial resolution of 1,280 × 800 pixels, at a frame rate of 50,000 ~frames~s$^{-1}$ and at an exposure time of 19 $\mu$s. The optical resolution is in the range of 112 to 116 pixels/mm. This configuration provided sufficient temporal and spatial resolution to easily identify and capture the entire necking and pinch-off dynamics for all the conditions explored in this work. 

\begin{table}[t]
\begin{tabular}{l c c c}
\hline
Liquid & $\mu$ (mPa\,s) \ \ & $\rho$ (kg/m$^{3}$) \ \ & $\sigma$ (mN/m) \ \\
\hline \hline
Water                           & \ \ \ \ \ 1.0     & 1,000 & 72.0 \\
90\% water / 10\% glycerin     & \ \ \ \ \ 1.5     & 1,005 & 71.5 \\
70\% water / 30\% glycerin     & \ \ \ \ \ 2.8     & 1,033 & 69.2 \\
50\% water / 50\% glycerin     & \ \ \ \ \ 6.8     & 1,082 & 67.4 \\
40\% water / 60\% glycerin        & \ \ \ \ \  12.1 &   1,115  &  66.6 \\
30\% water / 70\% glycerin     & \ \ \ \ 32.9    & 1,156 & 65.8 \\
20\% water / 80\% glycerin        & \ \ \ \   56.5 &   1187  &   64.8 \\
10\% water / 90\% glycerin     & \ \ \ 225.1   & 1,206 & 63.8 \\
5\% water / 95\% glycerin      &  \ \ \  505.3 & 1,230  &  63.3 \\
Glycerin                       & 1,350.0  & 1,248 & 62.8 \\
90\% water / 10\% ethanol      & \ \ \ \ \ 1.4     & \ \ \ 951  & 50.4 \\
80\% water /  20\% ethanol           &  \ \ \ \ \ 1.9 & \ \ \ 938  &  38.5 \\
70\% water / 30\% ethanol      & \ \ \ \ \ 2.5     & \ \ \ 927  & 34.2 \\
50\% water / 50\% ethanol      & \ \ \ \ \ 2.5     & \ \ \ 886  & 28.6 \\
10\% water / 90\% ethanol      & \ \ \ \ \ 1.6     & \ \ \ 811  & 23.9 \\
30\% water / 70\% ethanol       &  \ \ \ \ \  2.3 & \ \ \  827  &  26.1 \\
5\% water / 95\% ethanol    &  \ \ \ \ \  1.5 & \ \ \  790  &  22.8 \\
Ethanol                        & \ \ \ \ \ 1.2     & \ \ \ 789  & 22.2 \\
Methanol                       &  \ \ \ \ \ 0.6 & \ \ \ 784  &  22.3 \\
Silicon oil (2 cSt)            & \ \ \ \ \ 1.8     & \ \ \ 900  & 17.7 \\
Silicon oil (5 cSt)               &  \ \ \ \ \  4.6 & \ \ \ 916  &  19.7 \\
Silicon oil (50 cSt)           & \ \ \ \ 48.6    & \ \ \ 960  & 20.8 \\
Silicon oil (350 cSt)          & \ \ \ 348.0   & \ \ \ 969  & 21.1 \\
Silicon oil (1000 cSt)         & \ \ \ 996.4   & \ \ \ 970  & 21.2 \\
\hline
\end{tabular}
\caption{\label{table1}
Physical properties of the tested liquids at $T = 20.0 \pm 1.0\,^\circ$C.
Values of viscosity $\mu$, density $\rho$, and surface tension $\sigma$
used in this work.}
\end{table}

\subsection{Liquid materials and characterisation}

All the liquids used in our experiments are well-known Newtonians, and are listed in Table \ref{table1}. They consist of aqueous solutions of glycerin and ethanol, pure methanol and silicon oils. Deionized water was sourced locally, from a IQ 7000 Milli-Q purification system (at 18.2 M$\Omega$.cm and 2.3 ppb TOC). (Pure) ethanol (TWL00000010683) and methanol (TWL0000001064) were obtained from Fisher Scientific, and glycerin was sourced from UK ONYX Ingredients (99.9 \% pure). Silicone oils were acquired from Sigma Aldrich; these are 5~cSt (317667), 50~cSt (378356), and 1000~cSt (378399). The 2~cst and 350~cSt silicone oils were obtained from Azeils (PMX-200).

\begin{figure*}
  \includegraphics[width=0.54\textwidth]{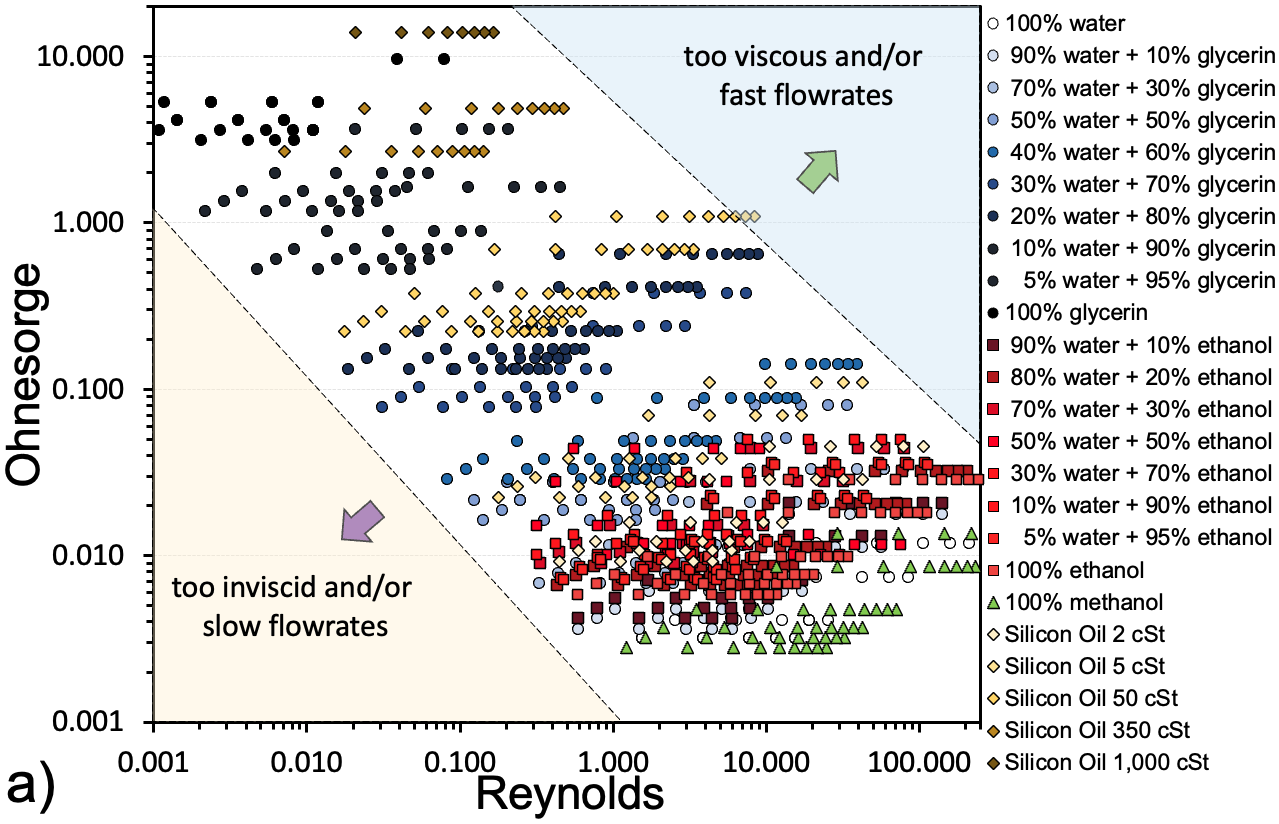}
    \includegraphics[width=0.45\textwidth]{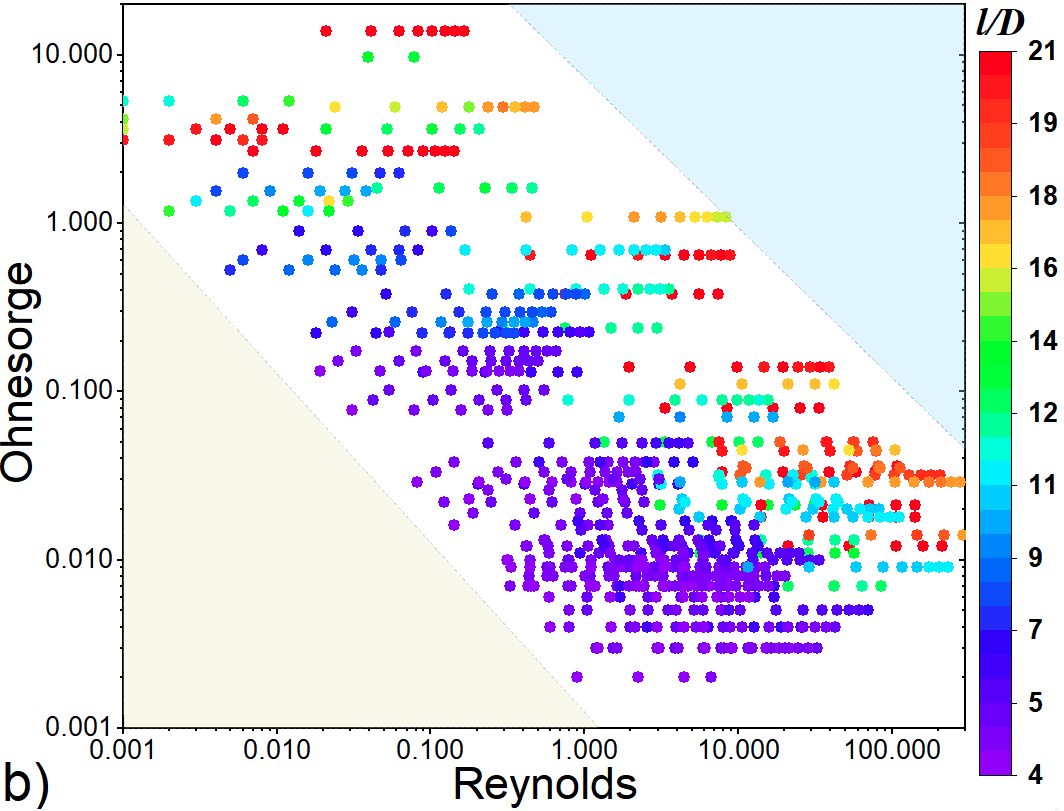}
    \caption{Data domains in terms of the Reynolds and Ohnesorge numbers. a) Parametric space of fluid properties and flow characteristics in terms of the liquid type and formulation; most liquids correspond to aqueous solutions of ethanol and glycerin. b) Parametric space in terms of the drop length to nozzle inner diameter ratio (l/D); as seen, the ratio is not a simple function of $Oh$ and $Re$ as many conditions lead to the same aspect ratio.} 
    \label{Domains}
\end{figure*}

The physical properties, i.e. density, viscosity, and surface tension, of the test liquids were characterised at (working) room temperature of (21.0 $\pm$ 1.0 Celsius). Density was determined gravimetrically by weighing a known volume of liquid using an analytical balance. Viscosity was measured on a rotational rheometer (Anton Paar MCR 102e) equipped with a cone–plate geometry at a constant shear rate of \(50\, \text{s}^{-1}\). Each sample was measured twenty times with the mean value taken as the working viscosity; the typical error in the viscosity measurement is $\delta \mu_{true} = \pm$ 0.10 mPa s. For liquids of low and moderate viscosity ( $\mu \lesssim 60~\mathrm{mPa\,s}$), the surface tension was evaluated via the maximum bubble pressure method using a SITA pro line T15 tensiometer at bubble lifetimes from 1.0~s to 15.0~s. Within this range, surface tension was found to be independent of surface age, with a standard error of $\delta \sigma_{true} = \pm 0.2$ mN/m. As noted in the introduction, bubble-pressure tensiometers are known to provide unreliable readings for highly viscous liquids due to viscous and hydrodynamic effects\cite{fainerman2004maximum, mysels1989some}. Accordingly, for glycerin and water mixtures of a concentration $\geq 80$~wt\% glycerin, and for 350~cSt and 1000~cSt silicone oils, surface tension values were taken from the manufacturers' data-sheets. The resulting liquid property data is presented in Table~\ref{table1}; we note that viscosity values span more than three orders of magnitude while surface tension triples in value. The domain of our data, parametrized in terms of $Re$ and $Oh$, can be found in Fig.~\ref{Domains}a. As can be seen there, our conditions cover four orders of magnitude in the Ohnesorge number and five in Reynolds, and include simple dripping, and jetting as defined in previous works \cite{ambravaneswaran2004dripping}.

\subsection{Image Analysis}
\textcolor{black}{In our method, high-speed images were analyzed to identify the frame at which full break-up occurred. The preceding frame (within 20~$\mu$s, as limited by the frame rat) was then selected, recorded, and analyzed. This process was repeated for each experiment. We carried out 840 experiments with varying conditions}, each producing an image (snapshot) showing the break-up of the droplet. Each image was processed by a custom MATLAB code to obtain the droplet profile and its contour. The code works by first subtracting a pre-recorded background (image without a droplet, just the background) from the droplet snapshot to enhance the interfacial contrast. The code then employs a global threshold function (Otsu's method) to obtain a binary image\cite{otsu1975threshold}. This is followed by steps to remove image artefacts and holes. The largest connected region in the binary image is then identified as the droplet, and its exterior boundary is extracted as the raw contour. Finally, the boundary is smoothed and resampled uniformly. \textcolor{black}{The needle outer diameter was used as size reference (pixels to micrometers) to determine lengths, such as the droplet diameter $D_d$ at pinch-off, and the drop length $l$ (the axial distance from the nozzle exit plane to the lowest point of the drop (apex) along the centerline), this for each experiment.} Figure~\ref{fig:liquids_and_processing} illustrates some of the steps of the image analysis, and the droplet shape at the point of break-up, for: a) pure glycerin; and b) water. These boundary-only silhouettes constitute the sole image inputs for our computational methods.

\begin{figure*}[t]
  a) \includegraphics[width=0.6\textwidth]{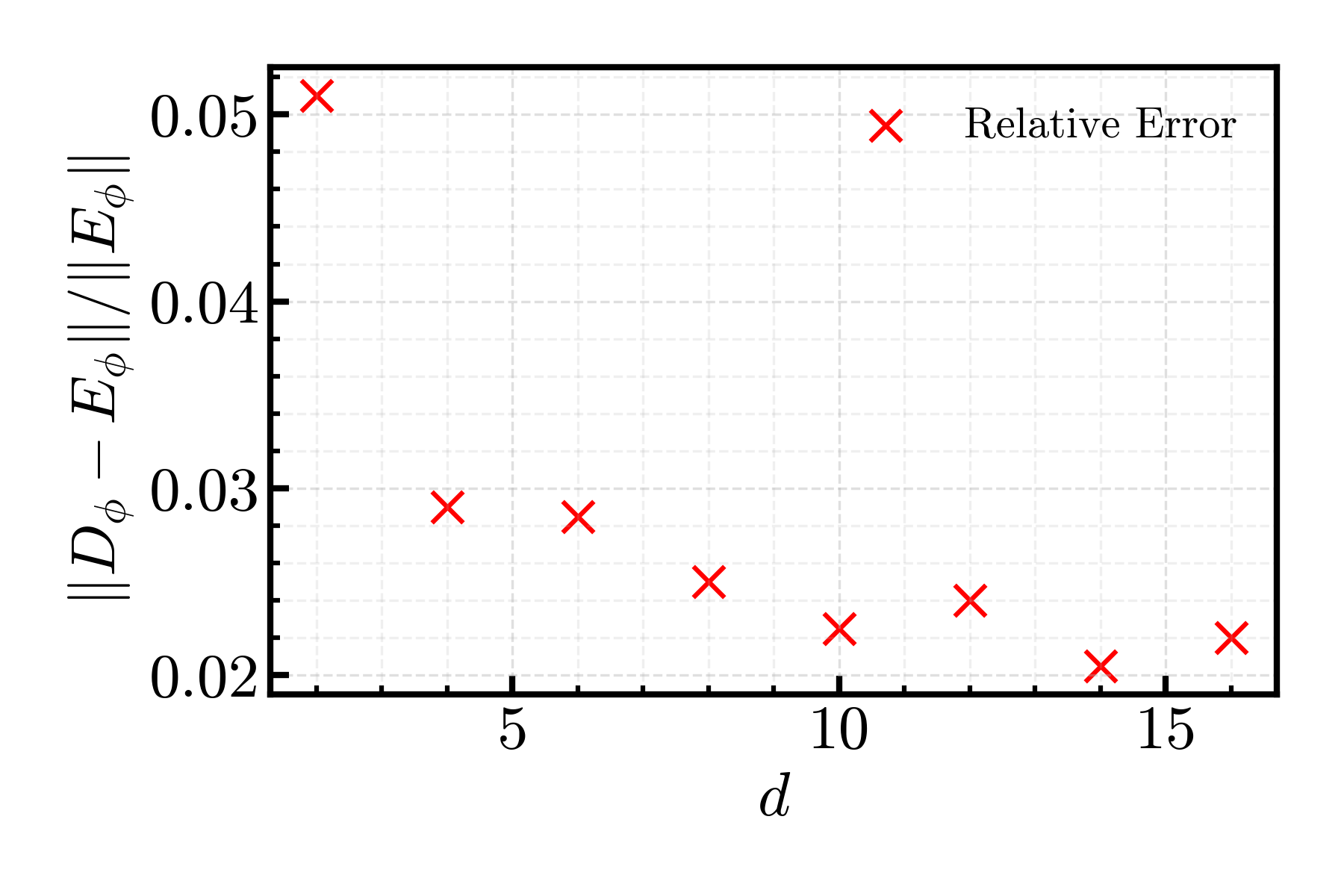} 
    b) \includegraphics[width=0.25\textwidth]{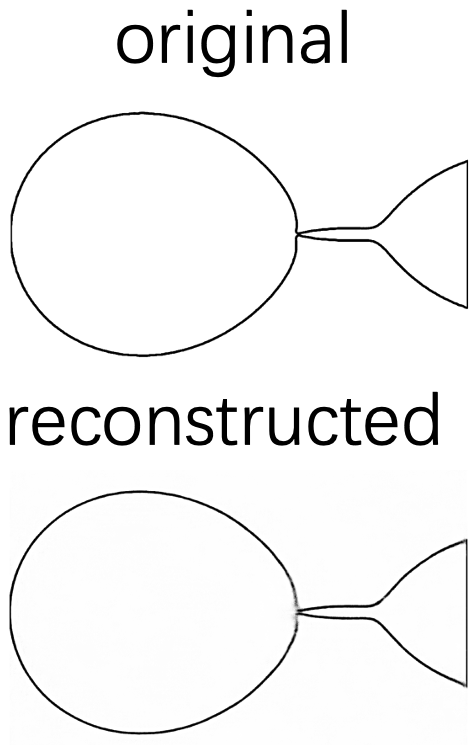}
    \caption{Autoencoder error analysis : (a) Relative reconstruction error $\|D_{\phi}-E_{\phi}\|/\|E_{\phi}\|$ on the validation set versus the latent dimension $d$ of the shape vector ($z\in\mathbb{R}^{d}$). As seen, the minimum error is found at $d=14$, which is the value used in our models. (b) Qualitative comparison of a representative boundary: ground truth (\emph{original}) vs. autoencoder output (\emph{reconstructed}). Here, $Re=7.943$ and $Oh=0.004$(90\% water + 10\% glycerin).}     \label{fig:autoencoder_error}
\end{figure*}

\section{Machine learning methodology} \label{sec:MachineLearningMethodology}

Supervised and unsupervised machine learning techniques were employed for the data-driven prediction of liquid properties and droplet shapes. The supervised methods aim  to either predict fluid properties based on physical inputs and latent shape features, or to reconstruct droplet shapes directly from physical parameters. In contrast, the unsupervised method focuses on identifying intrinsic groupings, i.e. clusters, of droplet characteristics based on experimental conditions and shape features.
\subsection{Extracting low-dimensional representation of boundary-only image with autoencoder}
A common foundation across both approaches is the use of a Convolutional Neural Network Autoencoder (CNN-AE) \cite{masci2011stacked} to extract low-dimensional representations of the droplet snapshots, in the form of a latent vector $\textbf{z}$. The structure of the autoencoder can be found in Table \ref{tab:autoencoder_arch} in Appendix \ref{app:autoencoder}. The primary motivation for employing a CNN is its ability to filter and de-noise raw image data while reducing the input dimension. These features are then used as model inputs in the supervised learning tasks, or as clustering coordinates in the unsupervised learning analyses. Our dataset is constructed from the droplet boundary-only images previously described. 
From the dataset of 840 samples we randomly selected 600 samples for training, 120 samples for validation, and 120 samples for the testing the autoencoder and the other methods. At each step, these groups of samples remained isolated from each other to avoid data leakage. The network was trained using the Adam optimizer \cite{kingma2014adam} with a learning rate of $2.5 \times 10^{-4}$. An early stopping criterion was applied: if the validation loss did not decrease for 30 consecutive epochs, training was terminated. Under these conditions, the training converged after 140 epochs with a batch size of 32.  Loss curves for the autoencoder are provided in Fig.\ref{fig:autoencoder_loss} in Appendix \ref{app:autoencoder}.

The model uses binary cross-entropy (BCE) loss, as the output is normalized. Mathematically, the autoencoder consists of an encoder $E_\theta$ and decoder $D_\phi$, parameterized by $\theta$ and $\phi$, respectively. Given an input image $x$, the BCE reconstruction loss is defined as

\begin{equation*}
\begin{split}
\mathcal{L}_{\text{BCE}}(x, D_\phi(E_\theta(x)))
    &= - \frac{1}{N} \sum_{i=1}^N \Big[
        x_i \log(D_\phi(E_\theta(x))_i) \\
    &\quad + (1 - x_i) \log(1 - D_\phi(E_\theta(x))_i)
    \Big].
\end{split}
\end{equation*}
\unskip

where $x$ is the ground truth image, $D_\phi(E_\theta(x))$ is the reconstructed image, both with pixel values in $[0,1]$, and $N$ is the total number of pixels. This loss encourages the model to accurately predict pixel intensities in the reconstructed image \cite{creswell2017denoisingautoencoderstrainedminimise}. 


The validation set was then used to evaluate the reconstruction performance of the autoencoder. Figure~\ref{fig:autoencoder_error}a shows the relative reconstruction error as a function of the latent dimension. Our results indicate that the relative error decreases rapidly as the latent dimension $d$ (the length of vector $\textbf{z}$) increases, and plateaus around $d = 14$, indicating that adding more dimensions does not improve accuracy. Consequently, in our methods, we set the autoencoder to produce a vector $\textbf{z}$ of size $d = 14$. We note that we have found a low-dimensional representation with a dimension in $\mathcal{O}(10^1)$, starting from an input dimension of $\mathcal{O}(10^5)$; thus a reduction of four orders of magnitude is achieved. A representative reconstruction example is found in Fig.~\ref{fig:autoencoder_error}b, contrasting the original contour and the reconstruction (with $d = 14$), and demonstrating that the autoencoder effectively captures the overall droplet geometry.

\subsection{Predicting liquid properties \& droplet shape with supervised learning}

In this section, we describe the supervised models used to infer either one or two fluid properties (viscosity and/or surface tension) from images, or to predict the droplet shape from a set of known physical parameters. All supervised learning models were implemented using two types of algorithms: multilayer perceptron (MLP) and gradient boosting decision trees (XGBoost)~\citep{Chen_2016}, which differ in their learning paradigms. The four XGBoost tasks, or models, considered are summarized in Fig. \ref{fig:supervised_diagram_xgb}, and details the MLP model are found in Appendix \ref{app:supervised}. Neural networks learn smooth continuous mappings, while XGBoost captures nonlinear feature interactions through ensembles of decision trees—allowing a complementary comparison between deep and tree-based methods. Models 1–3 combine the autoencoder latent vector $\textbf{z}$ with physical parameters (nozzle diameter $D$, flow rate $Q$, density $\rho$, viscosity $\mu$, and surface tension $\sigma$) to predict fluid properties. Model 4 performs the inverse mapping, predicting droplet shape from the same physical parameters, thereby establishing a bidirectional link between geometry and material behaviour. 
In addition, the same train, validation, and test split as used for the autoencoder was used here, with the same strict separation of these sets, in order to prevent data leakage. Each model was trained using the Adam optimizer with an initial learning rate of $10^{-3}$ and the mean-squared-error (MSE) loss function,
\begin{equation*}
\mathcal{L}_{\mathrm{MSE}}
    = \frac{1}{N} \sum_{i=1}^{N} \| \hat{y}_i - y_i \|_2^2.
\end{equation*}
where $N$ denotes the number of samples in a batch. 
A batch size of 32 was employed, and a StepLR scheduler was applied with a step size of 100 epochs and a decay factor of $0.8$. Early stopping with a patience of 100 epochs was applied based on validation loss, and the model weights corresponding to the lowest validation loss were restored for testing.
For targets with large dynamic range (e.g., viscosity), the outputs were log-transformed during training to improve numerical stability, and then mapped back to the physical scale for evaluation.

\textcolor{black}{The performance of the models was assessed by the coefficient of determination $R^2$, defined as
\begin{equation*}
R^2 = 1 - \frac{\sum_{i=1}^{n} (y_i - \hat{y}_i)^2}{\sum_{i=1}^{n} (y_i - \bar{y})^2},
\end{equation*}
where $y_i$ denotes the true (experimentally obtained) values, $\hat{y}_i$ the (ML-obtained) predicted values, and $\bar{y}$ the mean of the true values.} 


\begin{figure*}[t]
   \includegraphics[height= 8 cm]{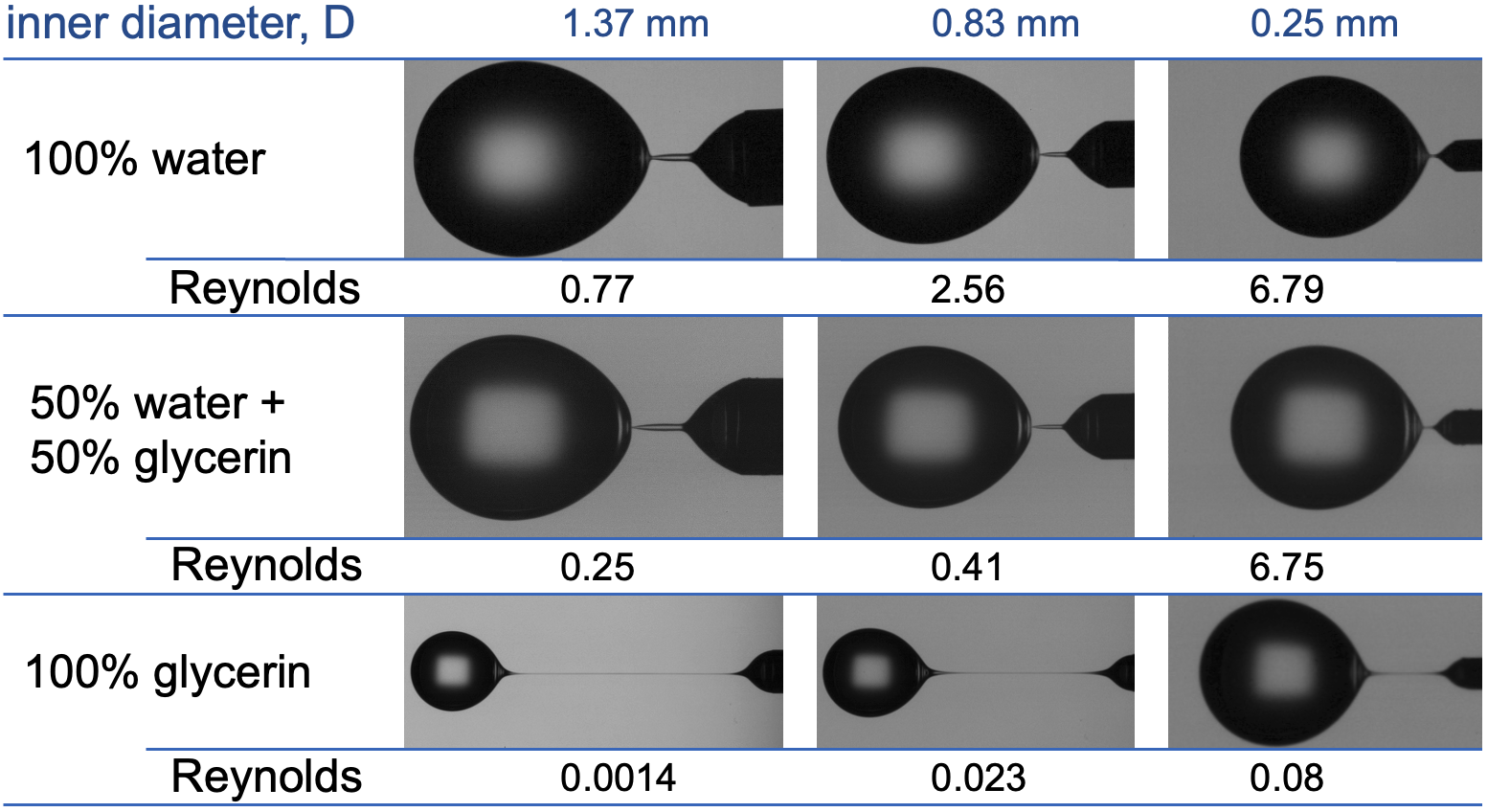}
    \caption{Examples of breaking up droplets at various conditions of Reynolds numbers for different liquids. As expected, pure water shows self-similarity near the pinch-off point and pure glycerin presents long filaments that break-up from the middle. Snapshots of pure glycerin are not to scale.} 
    \label{figure6}  
\end{figure*}

\subsection{Uncovering natural groupings in droplet behaviour with unsupervised learning} \label{sec:UncoveringNaturalGroupings}

In addition, we have employed unsupervised learning to directly uncover naturally occurring group behaviour in droplet dynamics from the observed shapes and imposed flow conditions, without requiring prior knowledge of the underlying fluid properties.

This work is motivated by experimental scenarios in which viscosity or surface tension are not readily measurable, yet droplet morphology and control parameters are accessible, e.g. in systems where liquid volumes are too small. 
\textcolor{black}{The autoencoder latent vector $\textbf{z} \in \mathbb{R}^{14}$ was} combined with two directly controlled experimental parameters, nozzle inner diameter and flow rate, for clustering. Notably, we deliberately excluded any information related to the fluid properties, as these quantities are often difficult to measure in practice. We applied both K-Means and Gaussian Mixture Models (GMM) with cluster numbers $K$, whose algorithm settings are provided in Appendix \ref{app:clustering_algorithms}. To identify optimal settings, we systematically tested $K \in \{2,7\}$. The choice of the upper bound $K=7$ reflects a practical consideration: while larger values of $K$ might further partition the data, excessively fine-grained clustering may produce groups which lack clear physical interpretation in the context of droplet dynamics. 
\textcolor{black}{After the evaluation of $K$  discussed in Appendix \ref{app:cluster_evaluation}, we adopt $K = 5$ in this work.}\

\begin{figure*}[t]
    \centering
    \includegraphics[width=\linewidth]{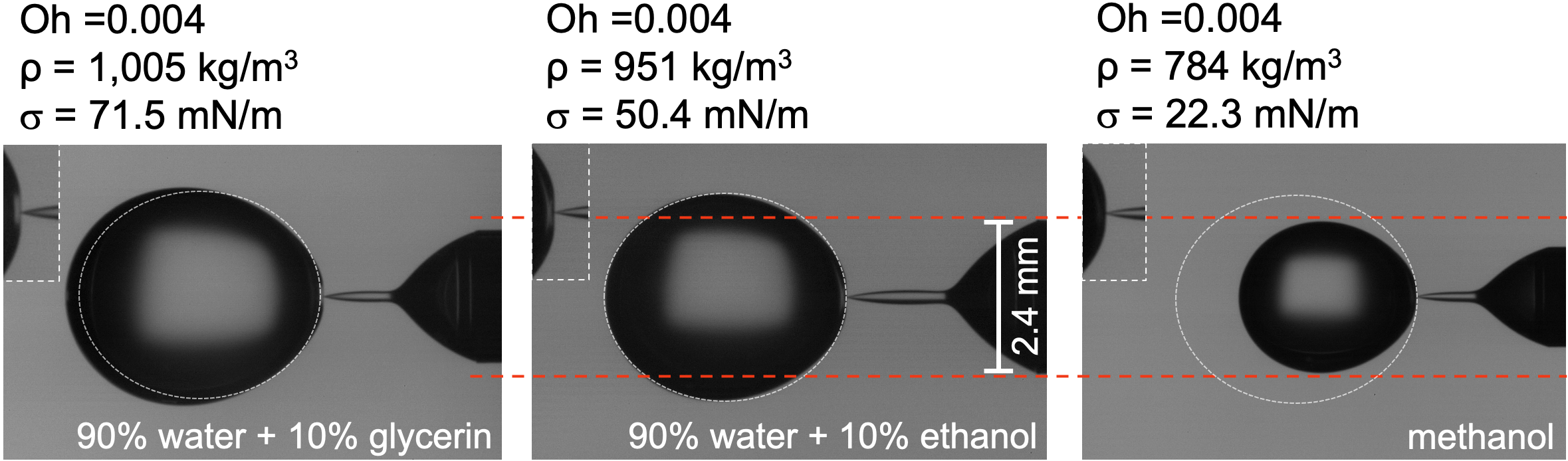}
    \caption{\color{black}{
    Break-up in the inertial-capillary regime for three different liquids at $Oh = 0.004$. The interface near the singularity exhibits apparent self-similar behaviour, as shown in the insets. However, the shape away from the location of the break-up differs significantly. The red dashed lines indicate the nozzle sizes, while the white dashed line compares the droplet size; the water and glycerin solution produces the largest droplet, whereas  methanol produces the smallest.}}
    \label{fig:similarity}
\end{figure*}

\section{Experimental observations} \label{sec:ExperimentalResults}

\textcolor{black}{Figure~\ref{figure6} shows example snapshots one frame prior to break-up} The images are organised by nozzle diameter $D$ and the liquid composition. As can be seen, water (a near inviscid fluid) displays a characteristic sharp neck that previous works have confirmed takes a 18 degree angle. In contrast, the 50\% mixture forms a longer, more gradually evolving neck, which indicates viscosity is playing a role in the dynamics. Interestingly, regardless of having very similar Reynolds numbers, pure water and 50\% glycerin (with $D=0.25$ mm) have visibly different shapes, with the glycerin solution having a larger bulb. Pure glycerin shows a slender, quasi–cylindrical long filament, that pinches off at the middle, consistent with a viscous–dominated dynamics. The Reynolds numbers listed beneath the panels increase as $D$ decreases, e.g. for water $Re=0.77 \rightarrow 2.56 \rightarrow 6.79$, and for glycerin $Re=0.001 \rightarrow 0.023 \rightarrow 0.080$). As can be seen, increasing viscosity (decreasing $Re$) shifts the minimum neck radius upstream and increases the break-up length to nozzle diameter aspect ratio.

The parameter map in Figure~\ref{Domains}a identifies different liquids within the $Re$–$Oh$ plane. As expected, methanol is found at the lowest $Oh$ values due to its low viscosity. In contrast, aqueous ethanols are found in the low–$Oh$, and moderate to high $Re$, region due to their low viscosities and surface tensions. Glycerin–water mixtures occupy intermediate $Oh$ values across a wide range of $Re$, with pure glycerin, and high–viscosity silicone oils, situating at high $Oh$ and low $Re$. The accessible domain of our setup is limited in the low left by very low viscosity and very slow flow rates, and in the upper right by the maximum pumping pressure, which limits the flow rates. We argue that our domain of experimental conditions is sufficiently extensive as it includes very low viscosity liquids (methanol and water), intermediate viscosities (solutions) and very high viscous fluids (silicon oil and glycerin), and very low surface tension (silicon oils) and high surface tension liquids (water). 

Figure~\ref{Domains}b provides an alternative perspective by color-mapping the break-up length-to-nozzle diameter ratio $L/D$ across the $Re$–$Oh$ space. As noted in the Introduction, the $L/D$ ratio has been used to determine whether a liquid filament will break up or not, and to distinguish dripping from jetting conditions \cite{ambravaneswaran2004dripping,notz2004dynamics}. From our results (the colour redundancy in Figure~\ref{Domains}b), various $Re$ and $Oh$ conditions lead to the same dimensionless break-up length; it is hence clear that the aspect ratio alone is insufficient to predict liquid properties, or a flow characteristic.

\textcolor{black}{Figure~\ref{fig:similarity} shows the shape of the interface for three breaking up droplets with the characteristic shallow angle, surface overturning, suggesting its route to self-similarity at the the singularity for the near-inviscid case. Here, at Oh  = 0.004, these characteristics are observed only near the location of the pinch-off as shown by \cite{day1998self}, but importantly these features do not extend to the whole fluid interface. Away from the capillary singularity region, differences in the relevant dimensionless ratios lead to distinct global droplet geometries and temporal evolutions. As seen in Figure~\ref{fig:similarity}, the interface near the (differently sized) nozzles, and the droplet sizes and shapes are very different. These larger-scale features provide the information used by our machine learning models to distinguish between different fluids and flow conditions, regardless of the apparent self-similar shapes near the point of break-up. We note that in the experiments seen in Figure~\ref{fig:similarity}, density varies from 1,005 to 784 kg/m$^3$, surface tension ranges from 22.2 to 71.5 mN/m, and viscosity is in the range of 0.6 to  1.5 mPa s.}

\bigskip

\begin{figure*}[t]
\vspace{3mm}
\centering
\setlength{\unitlength}{1cm}

\begin{picture}(8,5)
\put(0,0){\includegraphics[width=0.475\linewidth]{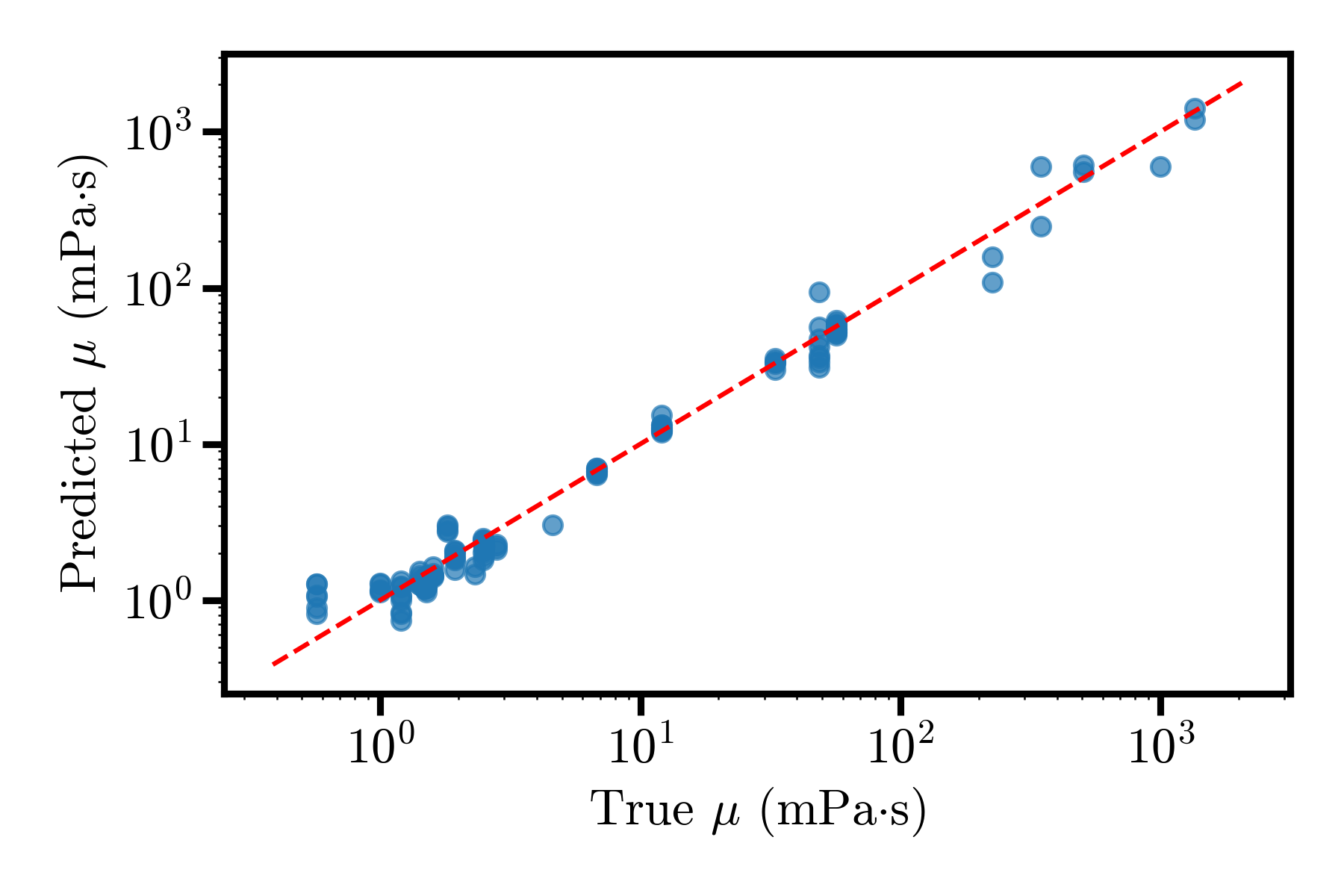}}
\put(1.6,4.8){\scriptsize $\mu_{\mathrm{pred}}=(0.91\pm0.02)\,\mu_{\mathrm{true}}+(2\pm4)$}
\put(2.5,4.4){\scriptsize $R^2 = 0.9428$}
\put(0.0,0.5){\scriptsize \textbf{(a)}}
\end{picture}
\hfill
\begin{picture}(8,5)
\put(0,0){\includegraphics[width=0.475\linewidth]{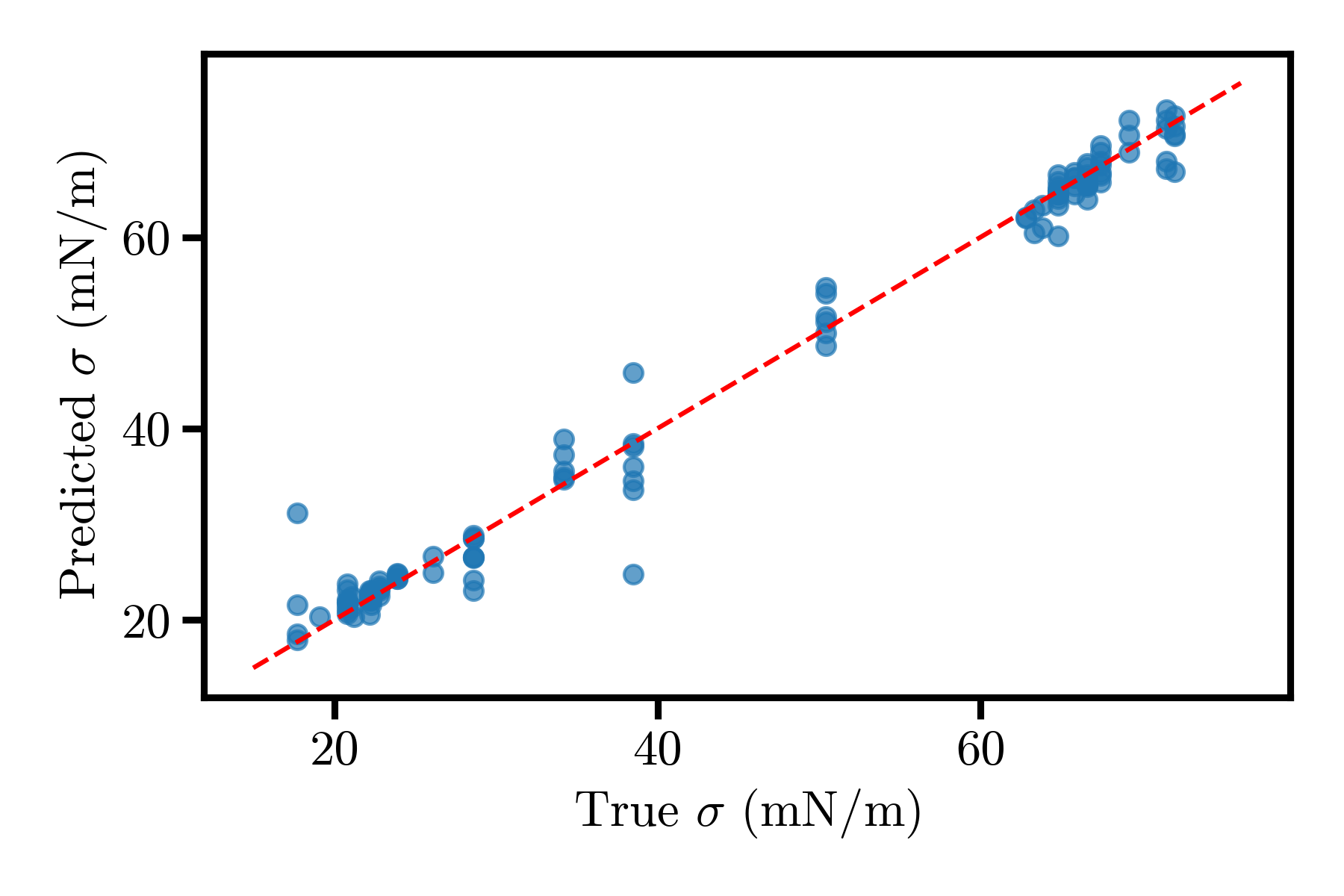}}
\put(1.5,4.8){\scriptsize $\sigma_{\mathrm{pred}}=(0.976\pm0.011)\,\sigma_{\mathrm{true}}+(1.0\pm0.6)$}
\put(2.5,4.4){\scriptsize $R^2 = 0.9843$}
\put(0.0,0.5){\scriptsize \textbf{(b)}}
\end{picture}

\vspace{3mm}

\begin{picture}(8,5)
\put(0,0){\includegraphics[width=0.48\linewidth]{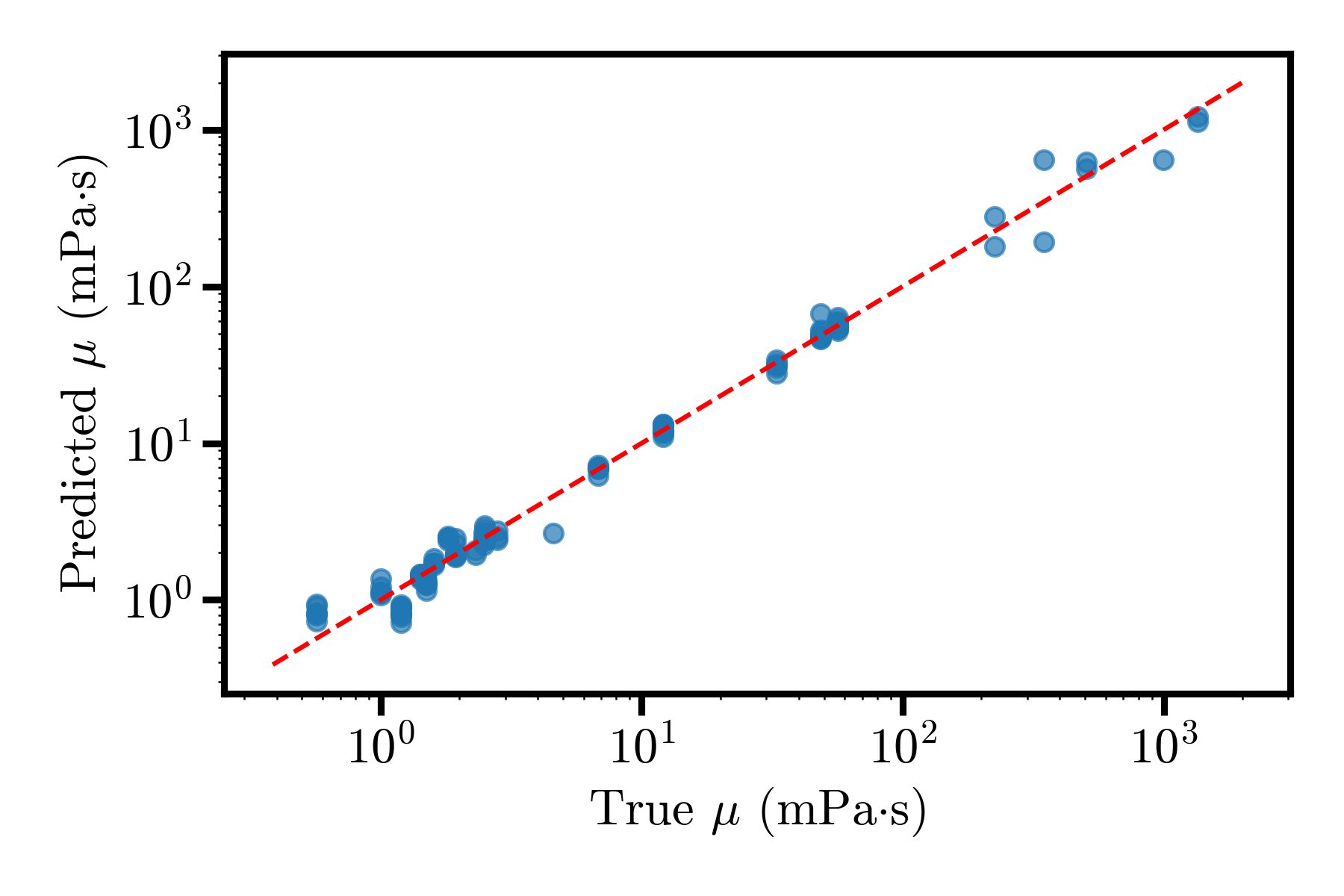}}
\put(1.6,4.8){\scriptsize $\mu_{\mathrm{pred}}=(0.87\pm0.02)\,\mu_{\mathrm{true}}+(5\pm4)$}
\put(2.5,4.4){\scriptsize $R^2 = 0.9361$}
\put(0.0,0.5){\scriptsize \textbf{(c)}}
\end{picture}
\hfill
\begin{picture}(8,5)
\put(0,0){\includegraphics[width=0.48\linewidth]{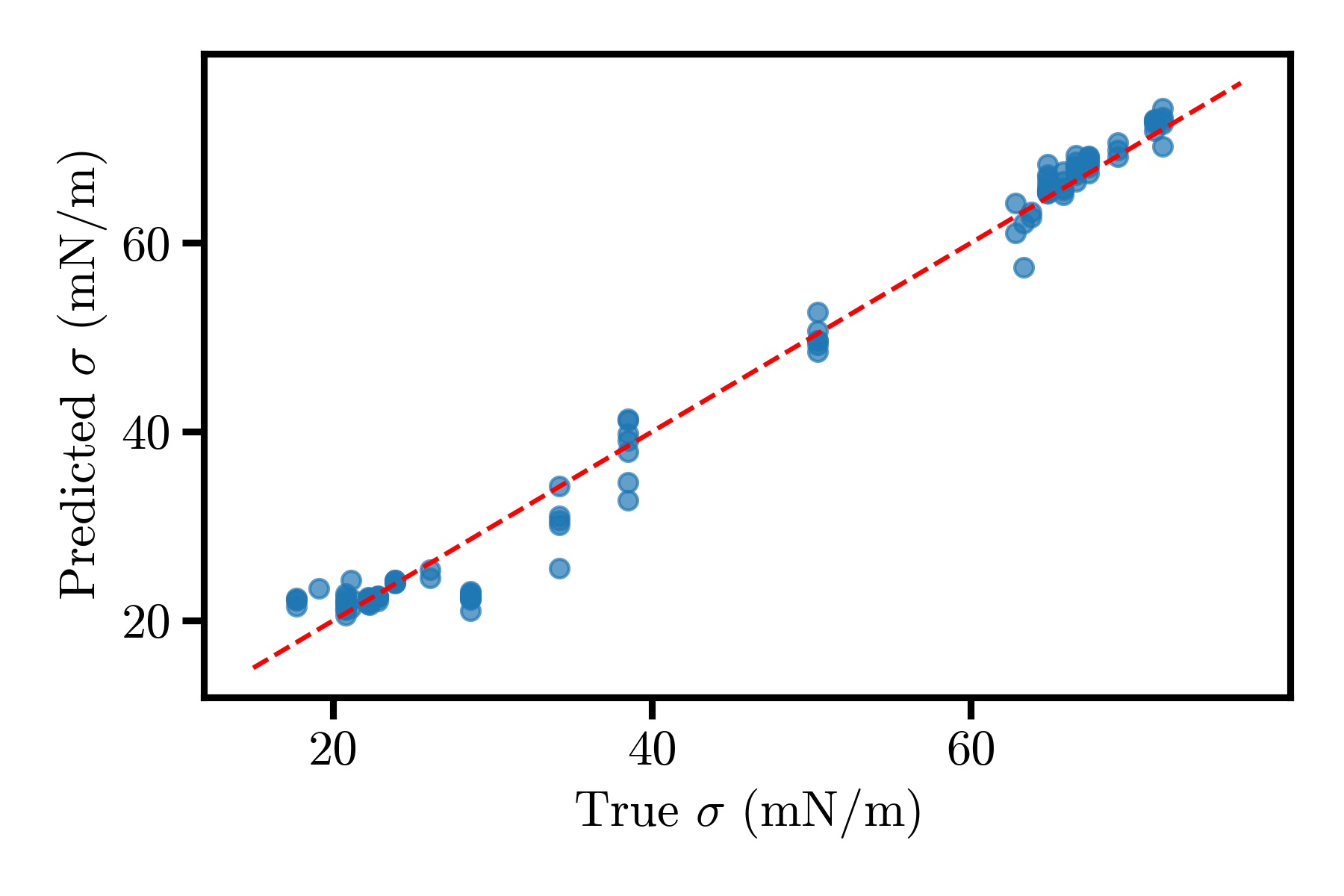}}
\put(1.5,4.8){\scriptsize $\sigma_{\mathrm{pred}}=(1.022\pm0.011)\,\sigma_{\mathrm{true}}-(1.0\pm0.5)$}
\put(2.5,4.4){\scriptsize $R^2 = 0.9854$}
\put(0.0,0.5){\scriptsize \textbf{(d)}}
\end{picture}

\caption{
Predicted vs.\ true values on the test set for Models~1--3 (MLP). 
Model~1 (a) predicts viscosity, Model~2 (b) predicts surface tension, 
and Model~3 predicts both viscosity (c) and surface tension (d).
}
\label{fig:combined_models}
\end{figure*}

\section{Machine learning results}

\subsection{Supervised learning results}

\begin{figure*}[t]
\vspace{3mm}
\centering
\setlength{\unitlength}{1cm}

\begin{picture}(8,5)
\put(0,0){\includegraphics[width=0.475\linewidth]{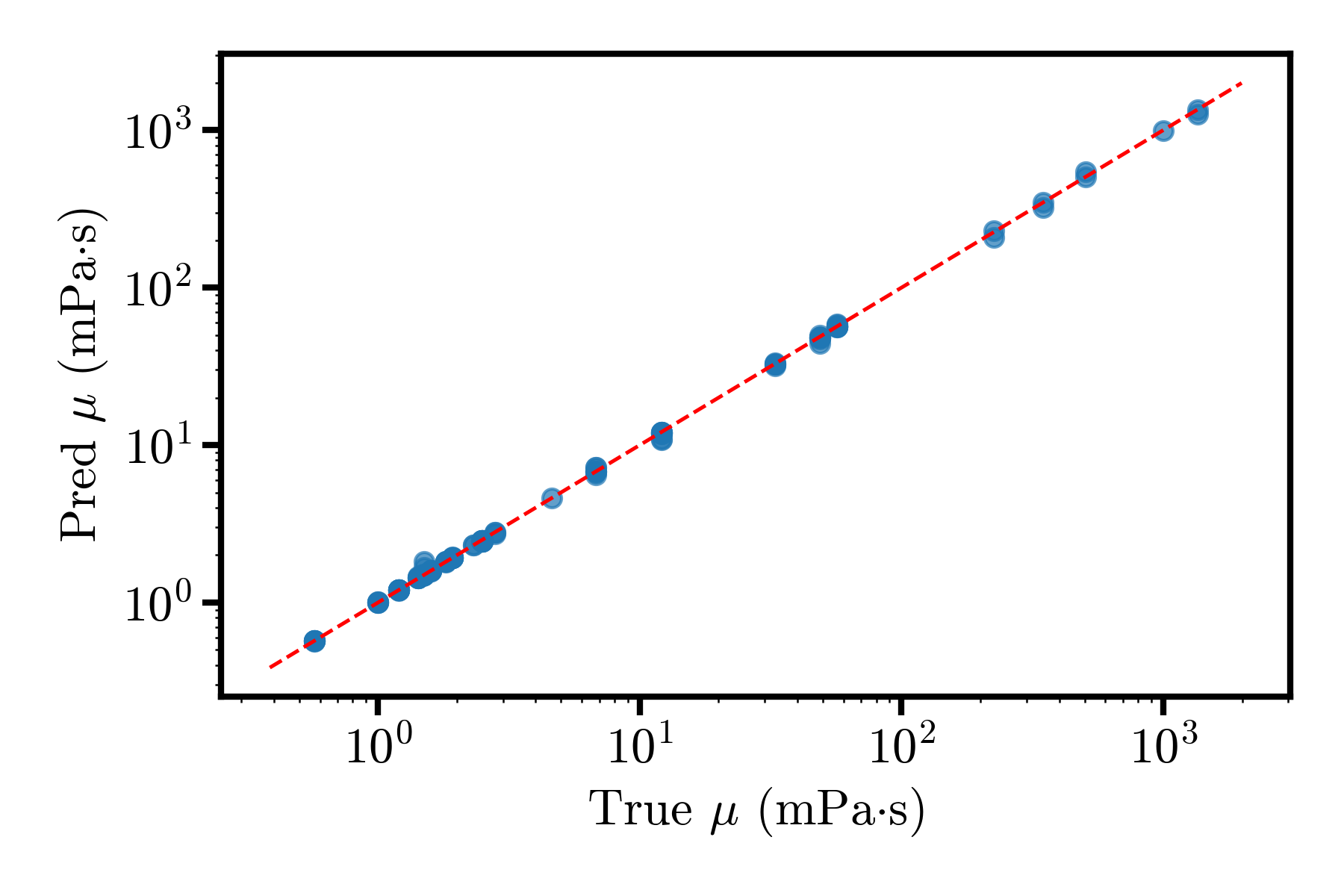}}
\put(1.6,4.8){\scriptsize $\mu_{\mathrm{pred}}=(0.977\pm0.004)\,\mu_{\mathrm{true}}+(0.5\pm0.8)$}
\put(2.5,4.4){\scriptsize $R^2 = 0.9978$}
\put(0.0,0.5){\scriptsize \textbf{(a)}}
\end{picture}
\hfill
\begin{picture}(8,5)
\put(0,0){\includegraphics[width=0.475\linewidth]{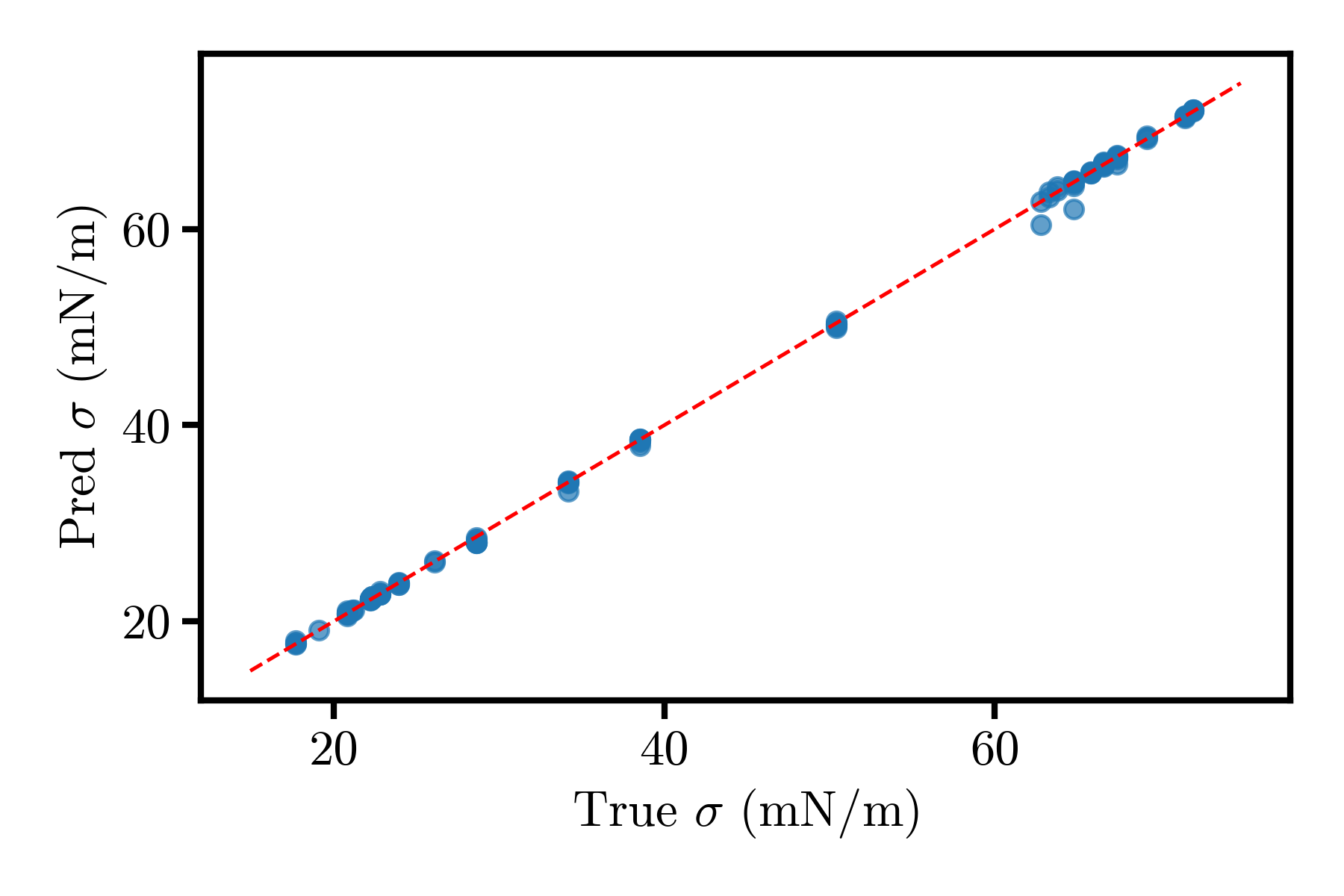}}
\put(1.5,4.8){\scriptsize $\sigma_{\mathrm{pred}}=(0.999\pm0.002)\,\sigma_{\mathrm{true}}-(0.05\pm0.09)$}
\put(2.5,4.4){\scriptsize $R^2 = 0.9996$}
\put(0.0,0.5){\scriptsize \textbf{(b)}}
\end{picture}

\vspace{3mm}

\begin{picture}(8,5)
\put(0,0){\includegraphics[width=0.48\linewidth]{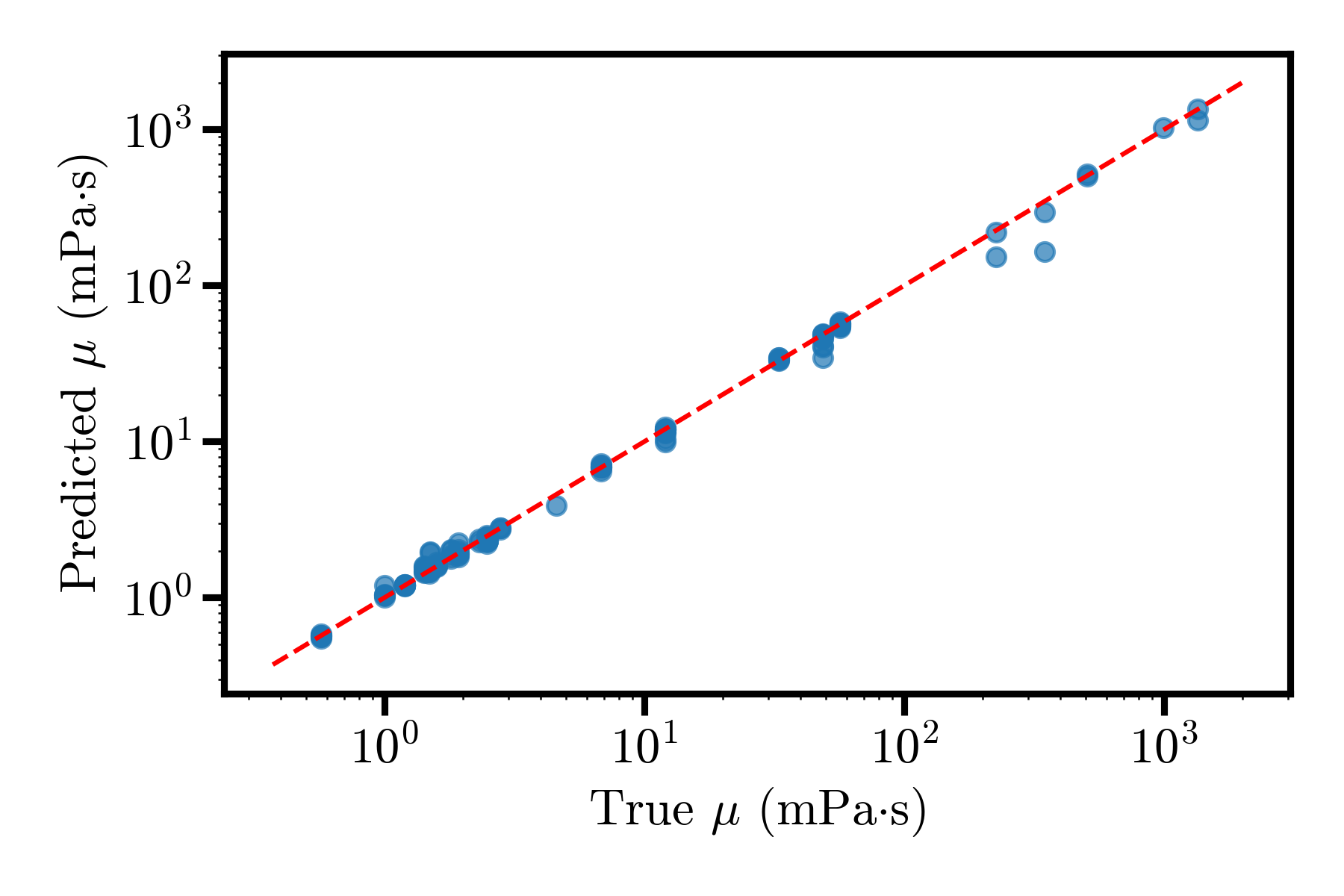}}
\put(1.6,4.8){\scriptsize $\mu_{\mathrm{pred}}=(0.937\pm0.010)\,\mu_{\mathrm{true}}-(0.5\pm2.2)$}
\put(2.5,4.4){\scriptsize $R^2 = 0.9831$}
\put(0.0,0.5){\scriptsize \textbf{(c)}}
\end{picture}
\hfill
\begin{picture}(8,5)
\put(0,0){\includegraphics[width=0.48\linewidth]{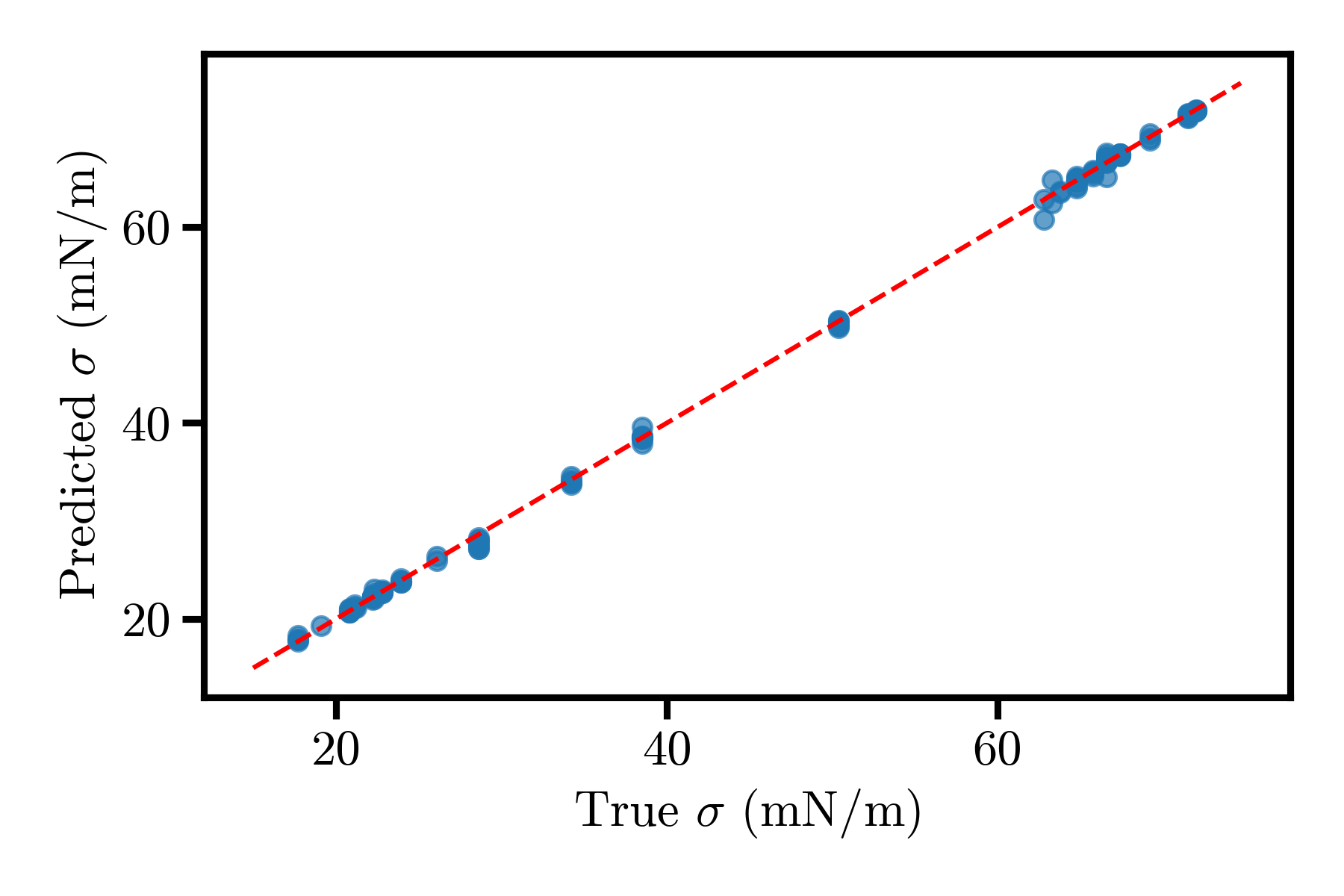}}
\put(1.5,4.8){\scriptsize $\sigma_{\mathrm{pred}}=(0.999\pm0.002)\,\sigma_{\mathrm{true}}-(0.05\pm0.10)$}
\put(2.5,4.4){\scriptsize $R^2 = 0.9995$}
\put(0.0,0.5){\scriptsize \textbf{(d)}}
\end{picture}

\caption{
Predicted vs.\ true values on the test set for Models~1--3 (XGBoost). 
Model~1 (a) predicts viscosity, Model~2 (b) predicts surface tension, 
and Model~3 predicts both viscosity (c) and surface tension (d).
}
\label{fig:xgb_results}
\end{figure*}

The four supervised models provide complementary perspectives on the extent to which fluid properties and droplet geometries can be inferred from a combination of physical parameters and learned shape representations. We evaluated the predictive performance of all four supervised-learning models on the test set, using the coefficient of determination ($R^2$) between predicted and true values as the main metric.

Model~1, implemented as a multilayer perceptron (MLP), predicts the viscosity from the other liquid properties, the flow characteristics, and the image-derived shape features. Figure~\ref{fig:combined_models}a uses a logarithmic scale and covers more than three orders of magnitude; we obtain $R^2 = 0.9428$. In contrast, with the same inputs, gradient boosting decision trees (XGBoost) reaches $R^2 = 0.9978$ (Fig.~\ref{fig:xgb_results}a), confirming that the viscosity signal is extremely well aligned with the latent geometry. We note that XGBoost is known to handle well small datasets. 

Model~2 addresses the complementary task of predicting surface tension from the same inputs as Model~1, but with surface tension removed from the inputs and viscosity provided instead. The MLP attains $R^2 = 0.9843$ (Fig.~\ref{fig:combined_models}b), while the XGBoost model slightly improves this to $R^2 = 0.9996$ (Fig.~\ref{fig:xgb_results}b), indicating that the latent vector indeed contains geometric markers that are controlled by capillarity.

Model~3 is the multitask setting, predicting both viscosity and surface tension together. Our MLP yields $R^2_\mu = 0.9361$ for viscosity and $R^2_\sigma = 0.9854$ for surface tension. The XGBoost version of the same task improves the numbers to $R^2_\mu = 0.9831$ and $R^2_\sigma = 0.9995$ (Fig.~\ref{fig:xgb_results}c and~\ref{fig:xgb_results}d). Despite having to predict two parameters simultaneously, the multi-task setting maintains essentially the same fidelity as the single-task models.

\begin{figure*}[t]
\centering
\vspace{3mm}
\setlength{\unitlength}{1cm}

\begin{picture}(8,5)
\put(0,0){\includegraphics[width=0.495\linewidth]{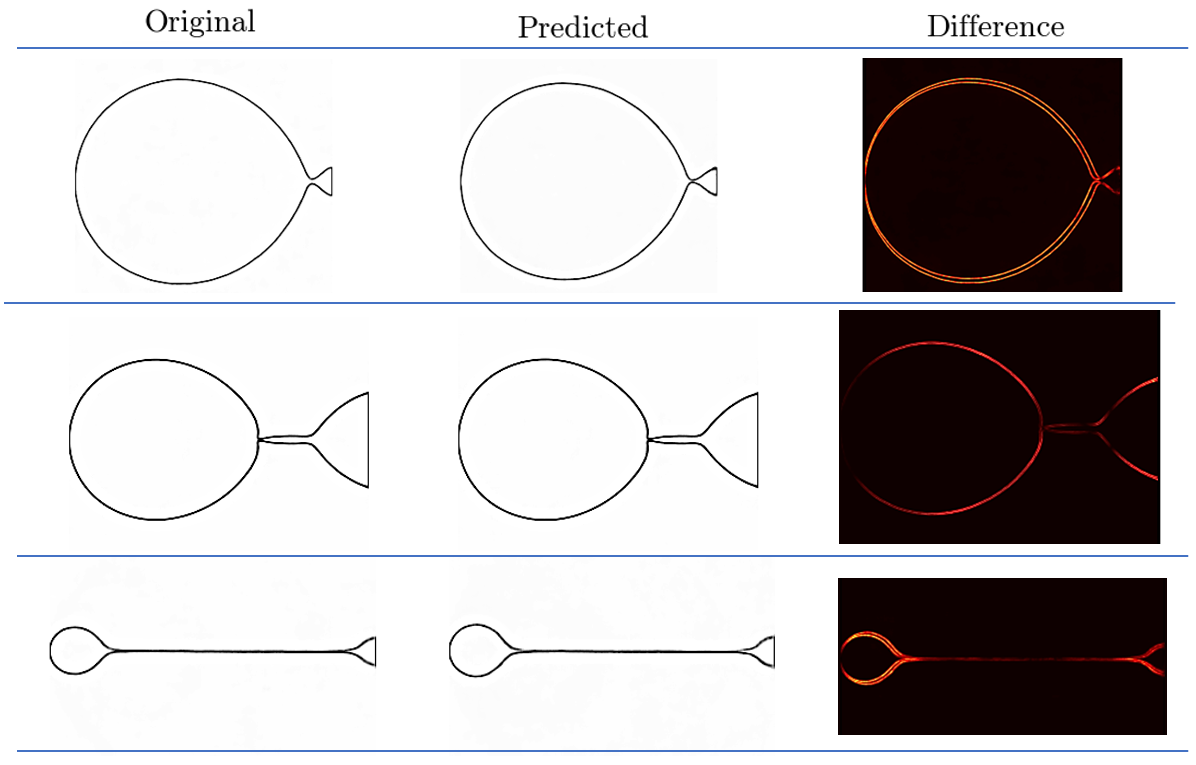}}
\put(0.1,-0.2){\scriptsize \textbf{(a)}}
\end{picture}
\begin{picture}(8,5)
\put(0,0){\includegraphics[width=0.495\linewidth]{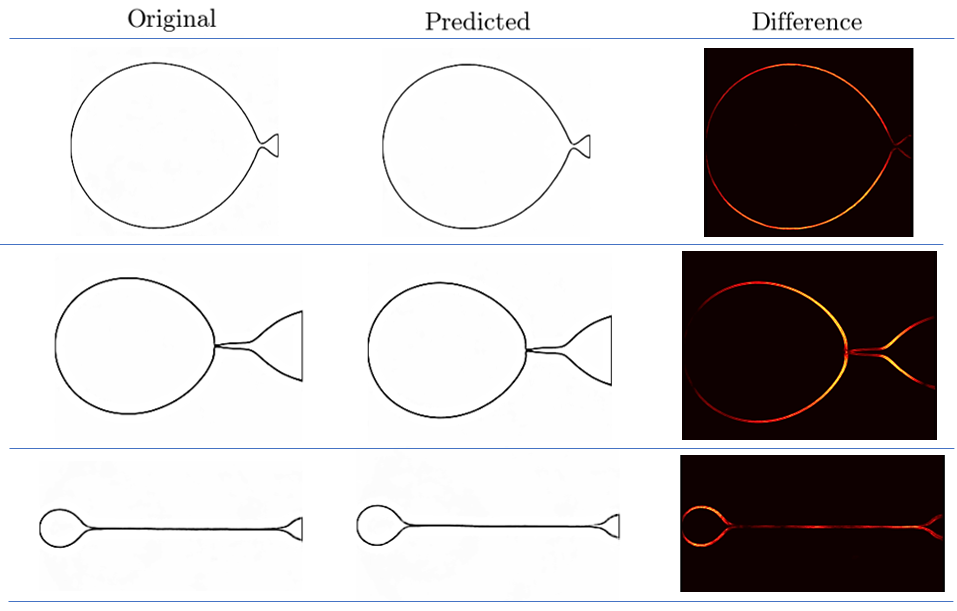}}
\put(0.1,-0.2){\scriptsize \textbf{(b)}}
\end{picture}

\begin{picture}(16,6)
\put(0,0){\includegraphics[width=0.95\linewidth]{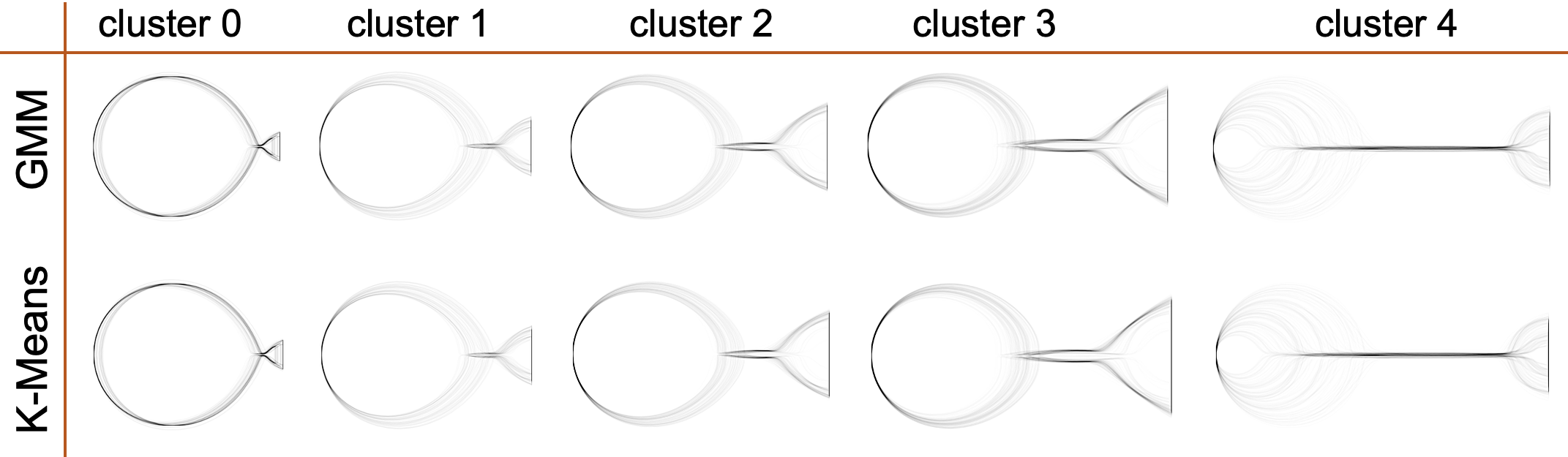}}
\put(-0.6,0.0){\scriptsize \textbf{(c)}}
\end{picture}

\caption{
Model 4 predictions: experimental vs.\ predicted droplet shapes for both
(a) MLP and (b) XGBoost variants.  
(c) Unsupervised learning results: average droplet shapes discovered by
GMM and $K$-Means clustering in latent space.
}
\label{fig:model4_results}
\end{figure*}

Model~4 differs from the others in that it predicts the \emph{shape} (the latent vector) directly from physical parameters, without any image as input. Using the MLP, we previously obtained $R^2 = 0.9145$ on the test set when comparing predicted latent vector. The XGBoost model reaches $R^2 = 0.9713$ in the normalized latent space. Because the decoder is a nonlinear map from latent coordinates to pixels, part of this advantage is smoothed out after decoding, but the high latent-space $R^2$ confirms that a large fraction of the image variability can be explained directly from the physical descriptors. Representative predictions are shown in Fig.~\ref{fig:model4_results}a and Fig.~\ref{fig:model4_results}b, we can see that both MLP and XGBoost models provide good prediction. Both models reproduce the global morphology reliably: the bulb radius, neck position, and overall aspect ratio match the ground truth. From the differences in the figures, it can be seen that MLP better maintained the continuity of the lines, but slightly underestimated the terminal bulbs, while XGBoost underestimated the thickness of the filament neck. 

We noted that, across all four supervised tasks, both the MLPs and the XGBoost models consistently achieve test-set $R^2 > 0.9$, and in the viscosity case the boosted model is effectively at the noise floor ($R^2 \approx 0.998$). This compares favourably with prior work \citep{gaikwad2022process,kim2022design}. Remarkably, the tree-based models (XGBoost) match or even slightly outperform the MLPs on several tasks. This is likely because the physical properties in our dataset, e.g. viscosity and surface tension, take only discrete values, determined by the limited categories of liquids used. \textcolor{black}{While the models presented here interpolate robustly within the experimentally sampled parameter space, the discrete set of fluid properties used in the training data may limit extrapolation beyond this range.} While the MLP attempts to learn a smooth, continuous mapping, XGBoost is naturally better suited for capturing piecewise-constant relationships arising from such discrete labels. The performance of all 4 models show that once a good latent vector is available, even a relatively shallow ensemble of boosted trees can exploit it extremely well.

\textcolor{black}{At first sight, the ability to infer fluid properties (such as surface tension $\sigma$ and viscosity $\mu$) from a single snapshot of a breaking droplet appears to contradict self-similarity and universality. In the asymptotic limit approaching pinch-off, the dynamics is governed by self-similar and universal solutions in which only ratios of material parameters enter (e.g., $\sigma/\rho$ for the inertial-capillary regime \cite{DayHinchLister1998}, and $\sigma/\mu$ for the inertia-capillary-viscous regime \cite{Eggers1993PRL}). In this strict limit, it seems that independent recovery of $\sigma$ and $\mu$ would therefore be impossible. However, a substantial body of experimental and theoretical work has established that it is extremely challenging to achieve the true asymptotic similarity regime during the dynamics of pinch-off. Instead, the approach to singularity is characterised by a complex sequence of transient regimes, slow convergence to similarity, and non-universal prefactors. Experiments and simulations have shown that neck thinning does not proceed via a single direct transition between canonical regimes (\cite{DayHinchLister1998, Eggers1993PRL, Papageorgiou1995POF}), but rather through multiple intermediate dynamical states that delay the onset of the final similarity regime \cite{BasaranPNAS2014}. Moreover, similarity solutions may exhibit oscillatory or slowly converging behaviour, arising from complex eigenvalue spectra and stability properties \cite{Dallaston2021}. These transient regimes reflect different balances of inertia, viscosity, and capillarity, and their occurrence depends sensitively on the material parameters. As a result, even when the system approaches a nominal similarity regime, the interface observed near pinch-off still bears information about its early dynamics. In addition, deviations from idealised similarity can arise when additional physical effects introduce additional length or time scales, such as heterogeneity or microstructure. These results highlight that the break-up dynamics is inherently path-dependent, and that the evolution of the interface (of the whole system, from nozzle, through to neck, to bottom of the drop) cannot, in general, be described by a single universal solution.}

\textcolor{black}{Therefore, the (whole) droplet morphology observed at the moment of break-up does not necessarily correspond to the universal similarity solution, but rather a finite-time snapshot of a trajectory through multiple dynamical regimes. Crucially, the sequence of regimes, the evolution of prefactors, and the spatial location of regime transitions depend separately on the fluid properties, as well as on flow conditions and geometry. The morphology therefore encodes the \emph{route to singularity}, rather than merely the final asymptotic scaling law. Attempts to infer surface tension by fitting pinch-off data to asymptotic scaling laws with fixed prefactors have been shown to produce significant errors when the system is not fully in the similarity regime \cite{Hauner2017}. In contrast, the present approach does not assume universality; instead, it leverages the non-universal, parameter-dependent morphology present in finite-time break-up dynamics.}

Within this framework, machine-learning models act as nonlinear feature extractors that map the high-dimensional geometric information contained in the shape of the interface to the underlying material parameters. The success of this approach therefore reflects not a violation of similarity theory but rather the fact that experimentally accessible pinch-off dynamics captures the regime in which independent information about $\sigma$, $\rho$, and $\mu$ (and other conditions) is preserved.

\subsection{Unsupervised learning results}
\label{sec:UnsupervisedLearningResults}

We employed unsupervised clustering on the feature representations extracted by the convolutional neural network to identify intrinsic groupings in the droplet dynamics, based on the droplet images in our validation dataset. From the behaviour of the Silhouette and DBI metrics across candidate values of $K$ (see Appendix \ref{app:cluster_evaluation}), we selected $K=5$ for both K-Means and GMM, corresponding to a maximum Silhouette Score. In the case of the DBI, extending $K$ beyond this value did not result in a large improvement, and, as previously noted, a smaller number of clusters is preferable for insight into the underlying physics.

We computed the average decoded shape of all droplets within a cluster to visualize the typical droplet morphology for that cluster. Figure~\ref{fig:model4_results}c shows these mean shapes for the GMM and K-Means partitions. Both methods readily and consistently identify Cluster~4 as a bulb attached to a very long filament, and Cluster~0 as a wide and rounded bulb coming from a-small-in-comparison nozzle (with no connecting filament). The other three clusters only differentiate in the relative size of the nozzle and its connecting filament. Cluster~1 consists of elongated droplets connected to a small nozzle through a short filament, Cluster~3 is a droplet connected, through a filament, to an equal-in-size nozzle, and Cluster~2 is a droplet connected to a intermediate-in-size filament and nozzle. The close agreement between the average shapes from the two methods confirms that both clustering approaches capture coherent and meaningful droplet shape features.

\begin{figure*}[t]
\vspace{15mm}
\centering
\setlength{\unitlength}{1cm}

\begin{picture}(8,5)
\put(0,0){\includegraphics[width=0.45\linewidth]{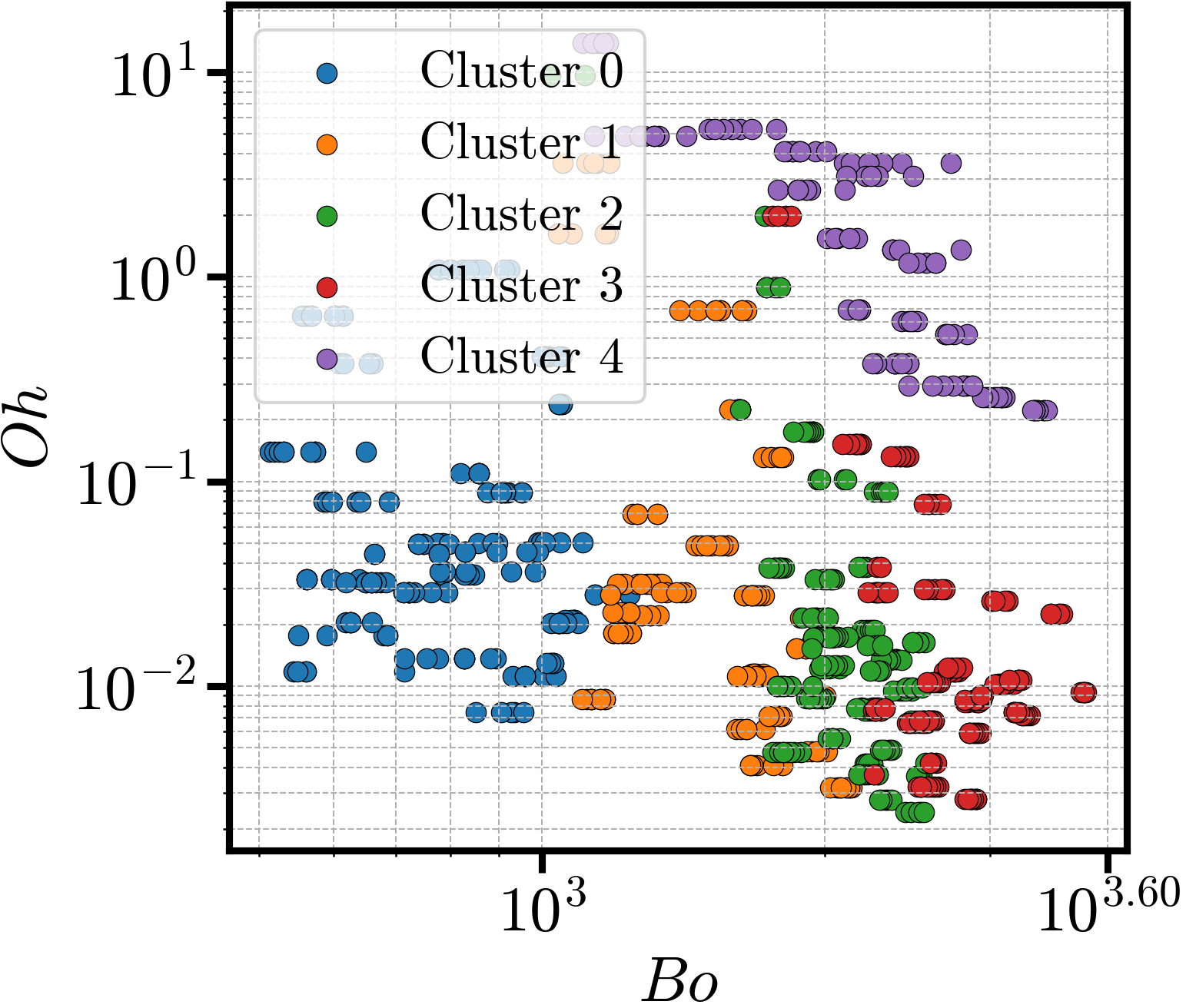}}
\put(0.3,0.4){\scriptsize \textbf{(a)}}
\end{picture}
\hfill
\begin{picture}(8,5)
\put(0,0){\includegraphics[width=0.45\linewidth]{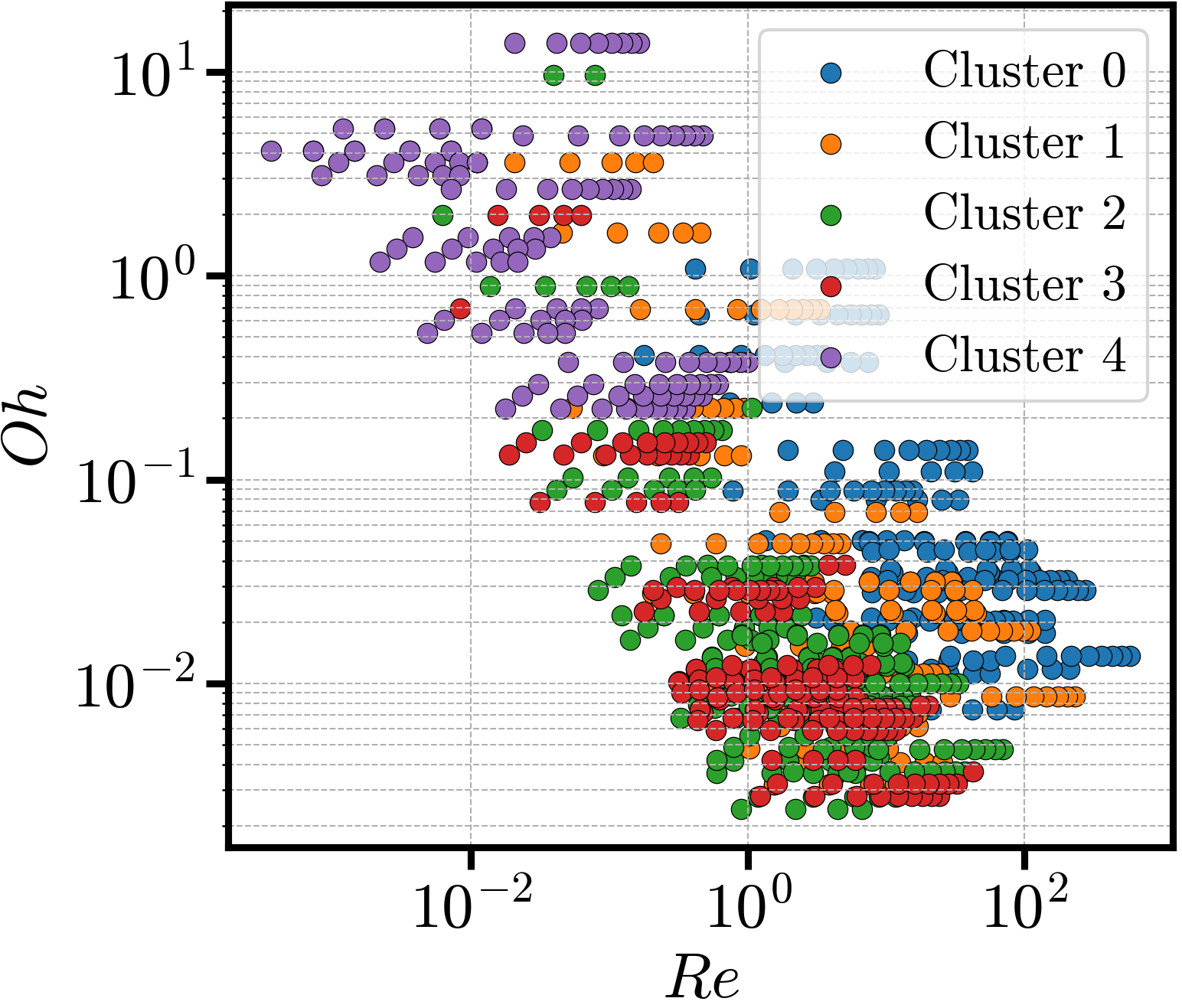}}
\put(0.3,0.4){\scriptsize \textbf{(b)}}
\end{picture}

\vspace{25mm}

\begin{picture}(8,5)
\put(0,0){\includegraphics[width=0.45\linewidth]{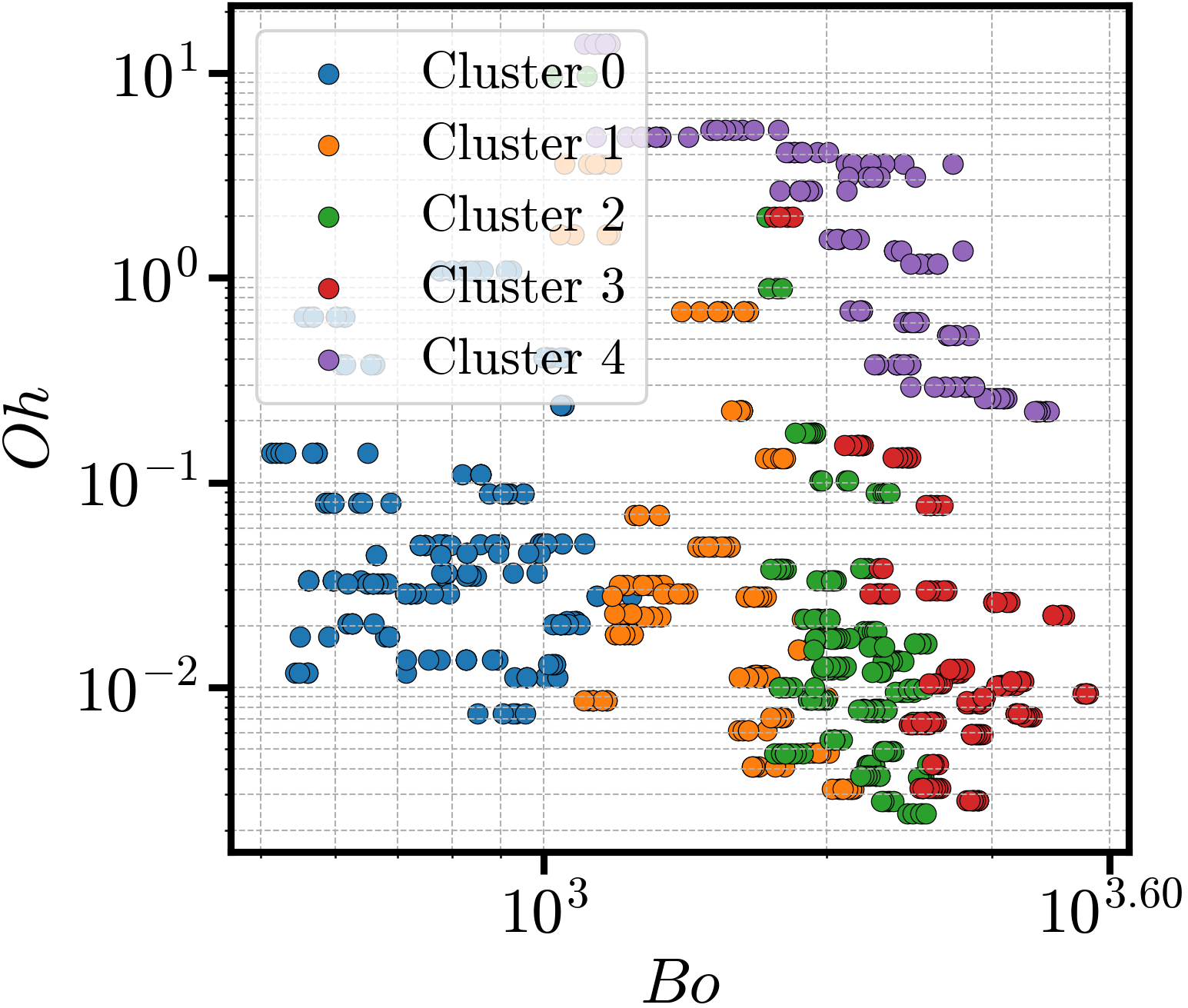}}
\put(0.3,0.4){\scriptsize \textbf{(c)}}
\end{picture}
\hfill
\begin{picture}(8,5)
\put(0,0){\includegraphics[width=0.45\linewidth]{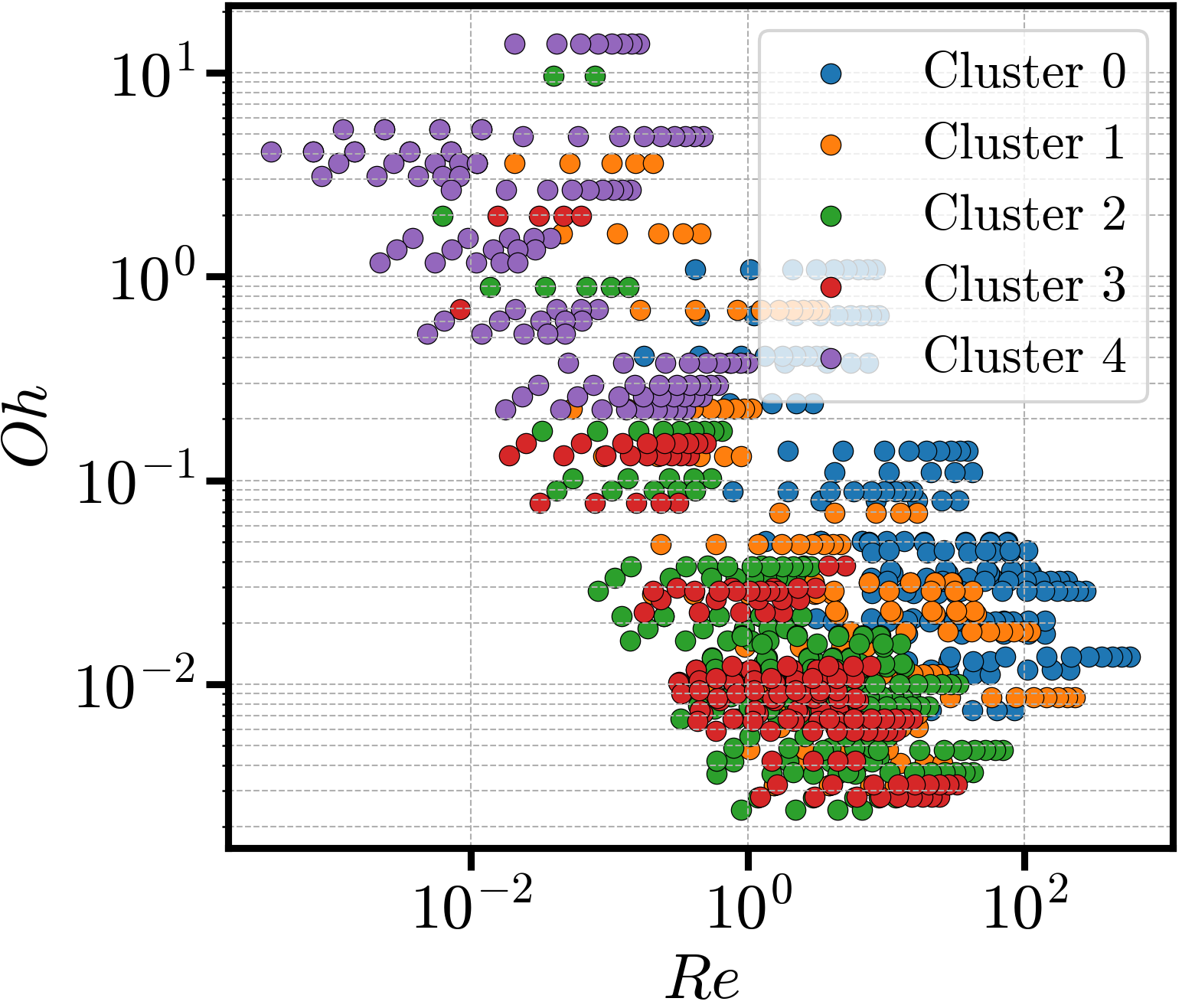}}
\put(0.3,0.4){\scriptsize \textbf{(d)}}
\end{picture}

\caption{
Clustering of droplet break-up regimes in the $Bo$–$Oh$ and $Re$–$Oh$ parameter spaces.
Panels (a) and (b) show $K$-means clustering, while panels (c) and (d) present the
Gaussian mixture model (GMM) results.
}
\label{fig:05}
\end{figure*}

A central question arising from the clustering is whether the unsupervised method effectively identifies, or differentiates, the underlying regimes and the various other factors influencing the dripping and break-up dynamics. Following previous works \cite{basaran2002small}, that have classified the liquid break-up dynamics in terms of scaling arguments, Figure~\ref{fig:05}b and Figure~\ref{fig:05}d show the cluster data projected onto the log-log ($Re$, $Oh$) plane for GMM and K-Means. Importantly, some clusters are readily physically interpretable. Decreasing $Oh$ indicates weakening viscous effects relative to inertial-capillary stresses, while increasing $Re$ reflects stronger inertia at fixed viscosity. Both methods clearly separate Cluster~4 and Cluster~0. Cluster~4 lies predominantly in the upper-left quadrant (high $Oh$, low $Re$), consistent with viscous-damped pinch-off with limited necking and slow recoil, and visually associated with very long filaments. In contrast, Cluster~0 concentrates in the lower-right region (low $Oh$, high $Re$), characteristic of inertial–capillary break-up producing small, nearly spherical droplets with weak viscous damping. Further analyses can be carried out by reflecting on the distribution of liquid formulations among the clusters, as summarized in the Table \ref{tab:kmeans_liquid_cluster} and \ref{tab:gmm_liquid_cluster} in Appendix \ref{app:cluster}. 
A clear pattern emerges from these analyses: high viscosity fluids, such as glycerin-rich solutions and high cSt silicone oils, are consistently grouped into Cluster~4 by both GMM and K-Means. These samples correspond to the most elongated droplets, supported by the high ratios $L_2/L_1$ and the mean shapes shown in Figure~\ref{fig:model4_results}c. In contrast, low-viscosity and low-surface-tension fluids are more frequently assigned to Cluster~0. These groups correspond to rounder, more stable droplets with minimal tailing.

Interestingly, Clusters 2 and 3 overlap in the low $Re$ and low $Oh$ quadrant, a region reserved for low viscosity liquids dripping at low flow rates. Furthermore, Cluster~1 is seen occupying the whole parametric space with no distinct regime, or liquid, or flow property in common. Intermediate formulations such as ethanol-rich fluids are more diffusely distributed across Clusters~2, 3 and 4, suggesting transitional regimes. The clustering results also show strong consistency between GMM and K-Means for the same liquids; this alignment reinforces that the latent feature space captures essential morphological cues that reflect fluid properties---even though clustering is entirely unsupervised and image-driven. 

The lack of a common denominator among data clusters 1, 2, and 3 suggests that a further factor, apart from the flow rate, the nozzle size, and the liquid characteristics, is playing a role in driving the dripping dynamics. Fortunately, dripping is well-known to be governed by a third dimensionless parameter, i.e. the Bond number \textcolor{black}{$Bo = {\Delta \rho g D_d^2}/{\sigma}$, where $\Delta \rho$ is the difference between liquid density and air density \cite{basaran2002small}. We note that, at the point of break-up, the nozzle size stops being the relevant characteristic length (as the flow is detached) and the droplet diameter $D_d$ is thus used in the Bond number. While $D_d$ is unknown a priori, unsupervised clustering does not require any input.} Figure~\ref{fig:05}a and~\ref{fig:05}c show the data and clustering in the $Bo$–$Oh$ space, where all five clusters are well separated, indicating distinct regimes governed by the relative importance of gravity ($Bo$) and viscosity ($Oh$).  Clusters~0-3 occupy a narrow range of $Oh$ but progressively shift toward larger $Bo$. This increase in $Bo$ indicates that these clusters primarily reflect the gradual strengthening of gravitational forces relative to capillarity, while viscous effects remain comparatively constant. Cluster~0 corresponds to the capillary-dominated regime, where droplets are nearly spherical and detach quasi-statically. The gravitational effect strengthens among Clusters~1-3; even though they have comparable viscosity, the increasing gravitation elongates the filaments. In contrast, Cluster~4 deviates markedly from this $Bo$-driven sequence; it spans to much higher $Oh$ values while maintaining moderate $Bo$, this corresponds to high viscous effects in which surface tension and gravity are both resisted by strong viscous damping.

As described in Section \ref{sec:UncoveringNaturalGroupings}, the clustering was performed on a feature vector that concatenates droplet-shape descriptors with the experimental settings (nozzle diameter and mean flow speed) and does not explicitly include fluid properties. The data separation in $Re$–$Oh$ and $Bo$–$Oh$ therefore suggests that these features implicitly encode the inertia–viscosity–gravity balance. Notably, the physical properties of the liquids were not input into the clustering. Thus the separation in Re-Oh and Bo-Oh space indicates that our method successfully uncovered the underlying physics linking the experimental settings to the observed droplets. The strong mapping between physical composition and cluster assignment provides further evidence that the unsupervised clusters are not only mathematically separable but also physically interpretable.

 \section{Uncertainty analysis}\label{sec:UncertaintyAnalysis}

\textcolor{black}{The supervised learning framework relies on experimentally measured fluid properties as ground truth, and the achievable prediction accuracy is therefore dependent on the uncertainty of these measurements. Despite this, the consistently high $R^2$ values suggest that the relationship between morphology and material properties remains robust, and the method provides a rapid characterization approach with an inherent trade-off between accuracy and simplicity. Experimental uncertainty in the training data introduces label noise, which can affect the learned mapping between morphology and fluid properties.}  Associating standard errors with machine learning predictions is not a common practice (to the best of our knowledge, largely unexplored). To address this issue, in this section, we propose a simple, yet formal, method to associate uncertainties to the predictions of the models. The idea is simple, and follows the standard method of contrasting predicted and true values. Figures \ref{fig:combined_models} and \ref{fig:xgb_results} show various predicted values plotted against known properties. As with any other method, we observe the prediction is not perfect, with the data showing some scatter around the known values, \textcolor{black}{i.e. the ground truth values}. When presented in this form, the predicted and true ground (experimentally measured) values can be linked by a linear fit, with a regression analysis able to provide us with average and error values for the slope and intercept. In brief, predicted and true values form the function

\begin{equation}
\zeta_{predicted} =  m \zeta_{true}+b, \label{PredictedTrue}
\end{equation}
where $\zeta_{predicted}$ is the predicted property, which in our case can be either the surface tension or the viscosity, $\zeta_{true}$ is the true \textcolor{black}{(measured, or ground truth)} value, $m$ is the slope, and $b$ is the value taken by the prediction at a condition where the true value is zero (or the point where the line crosses the y-axis). In a perfect scenario, $m$ would take a value of one, and $b$ the value of zero, i.e. every prediction matching the true value. However, our predictions are not perfect, $m$ is not one and $b$ is not zero. In fact, our true values are also not uncertainty-free as they were obtained experimentally with error, \textcolor{black}{this error is known as ground truth uncertainty in machine learning.} Propagation of uncertainty theory tells us that the error on the prediction (Eq. \ref{PredictedTrue}) is given by

\begin{equation}
\delta \zeta_{predicted} =  \sqrt{\left(\zeta_{true} \delta m \right)^2+ \left(m \delta \zeta_{true}\right)^2+\delta b^2 }, \label{ErrorPropagation}
\end{equation}
where $\delta \zeta_{predicted}$ is the error of the predicted value, $\delta \zeta_{true}$ is the experimental error of the true value \textcolor{black}{(ground truth uncertainty)}, $\delta m$ is the error of the slope, and $\delta b$ is the error of the y-axis crossing. Combining Eqns. \ref{PredictedTrue} and \ref{ErrorPropagation} we obtain:

\begin{equation} 
\delta \zeta_{predicted} = \sqrt{ \left(\frac{\zeta_{predicted} -b}{m}\right)^2 (\delta m)^2 + \delta b^2 + (m \delta \zeta_{true})^2}. \label{ErrorFinal}
\end{equation}

\textcolor{black}{As seen, the first two terms of this equation depend on the fitting coefficients ($m$ and $b$) and the third term depends on the ground truth uncertainty}. Applying this relationship to, for example, our viscosity predictions from the MLP model 1, Figure \ref{fig:combined_models}a, \textcolor{black}{i.e. $m=0.91$, $\delta m =0.02$, $b=2$ mPa s, $\delta b = 4$ mPa s and $\delta \mu_{true} = 0.1$ mPa s (ground truth uncertainty)}, we obtain that the error on the prediction is given by

\begin{equation*}
\begin{split}
\frac{\delta \mu_{\text{predicted}}^2}{\text{ (mPa s)}^2}
    &= 4\times10^{-4}\left(\mu_{\text{predicted}} - 2\  \right)^2 + 16+ 8\times10^{-3}. \\[0.3em]
\end{split}
\end{equation*}

\textcolor{black}{A simple analysis of this equation indicates that, for very low viscosities, the error on the prediction is largely dominated by $\delta b$, or the inaccuracy of the MLP method to predict very low viscosities. In fact, for viscosities near that of water, the error on the prediction is in the order of $\sim$ 4 mPa s. However, for higher viscosities, the fractional error quickly decreases (please look at the Appendix \ref{app:error}, Figure \ref{fig:error}a) and at medium viscosities, i.e. $\mu_{predicted}> 100$ mPa s, the  error associated to the MLP prediction only accounts to 5 percent. For even higher viscosities, i.e. $>500$ mPa s, the total error only amounts to 2 percent. Consequently, the MLP model is well suited to predict viscosities above 100 mPa s. We also note the third term of the equation, associated to the ground truth uncertainty, is never relevant to the overall uncertainty.}

In contrast, the error analyses of models 1 and 3 from XGBoost, e.g. Figure \ref{fig:xgb_results}a \textcolor{black}{($m=0.977$, $\delta m =0.004$, $b=0.5$ mPa s, $\delta b = 0.8$ mPa s and $\delta \mu_{true} = 0.1$ mPa s)} and Figure \ref{fig:xgb_results}d \textcolor{black}{($m=0.999$, $\delta m =0.002$, $b=0.05$ mN/m, $\delta b = 0.1$ mN/m s and $\delta \sigma_{true} = 0.2$ mN/m)}, produce much more accurate results. The error on the predicted viscosity is given by

\begin{equation*}
\begin{aligned}
\frac{\delta \mu_{\text{predicted}}^2}{\text{ (mPa s)}^2}
&= 1.7\times10^{-5}\left(\mu_{\text{predicted}} - 0.5\  \right)^2 
       + 0.64 + 0.01. \\[0.3em]
\end{aligned}
\end{equation*}
\textcolor{black}{Here, the analyses, (Figure \ref{fig:error}b in Appendix \ref{app:error}), also shows that the error associated to the XGBoost model at low viscosities is also largely associated to $\delta b$. However, the error on the prediction only accounts to $\mu_{\text{predicted}} = 0.8$ mPa s at low viscosities; five times smaller than in the MLP. In fact, XGBoost predicts within a single digit percentile viscosities in the range $\mu_{true} > 10$ mPa s, and the total fractional error is below 1 percent for viscosities $\mu_{true} > 100$ mPa s. Therefore, we conclude the MLP model is well suited to predict viscosities above 10 mPa s, and excellent for viscosities beyond 100 mPa s.}

\textcolor{black}{Finally, the error associated to predictions of surface tension by XGBoost is given by}
\begin{equation*}
\begin{aligned}
\frac{\delta \sigma_{\text{predicted}}^2}{(\text{mN/m})^2}
    &= 4\times10^{-6}(\sigma_{\text{predicted}} - 0.05)^2 + 0.01+0.04.\ \\
\end{aligned}
\end{equation*}
\textcolor{black}{Here, the dominating source of error is the ground truth uncertainty, i.e. the third factor in the equation is always the largest. In other words, the error associated with the experimental measurements exceeds the uncertainty introduced by the XGBoost model. We note $\delta \sigma_{predicted}$ stays below $1\%$ (Figure \ref{fig:error}c in Appendix \ref{app:error}), making the XGBoost model excellent for predicting surface tension.}

\section{Conclusions} \label{sec:Conclusions}

-------------

\textcolor{black}{
A central outcome of this work is that the interface geometry near pinch-off retains sufficient information to infer fluid parameters, such as surface tension  and viscosity, from a single experimental snapshot.
While asymptotic similarity predicts that, sufficiently close to singularity, the dynamics are governed by universal exponents depending only on ratios of material parameters,  finite-time breakup dynamics do not collapse instantaneously onto this state. Instead, the interface retains information about fluid properties and  flow conditions.
The break-up morphology can therefore be interpreted as encoding the \emph{route to singularity}, rather than merely the final similarity exponent. Classical similarity analysis attempts to eliminate non-universality; the  approach presented here exploits it. Machine-learning models can extract this high-dimensional morphological information without assuming universal prefactors. The morphology at breakup reflects the entire route to singularity, not only the final asymptotic scaling law.}

\textcolor{black}{To test this approach}, we have systematically analyzed the pinch-off dynamics of Newtonian liquids, including water, methanol, silicone oils, ethanol, and aqueous ethanol and glycerol–water mixtures, using high-speed imaging across 840 experimental conditions spanning 0.001 $<$ Re $<$ 200 and 0.01 $<$ Oh $<$ 20. The frame immediately preceding droplet break-up was captured to record the pendant drop shape at detachment. The contour of each droplet was extracted and standardized for subsequent machine learning analysis. 
Supervised regression models were used to predict viscosity and surface tension from these contours and known flow parameters. We further built an inverse model that predicts the shape of the drop neat the pinch-off time from physical inputs, and employed unsupervised clustering to reveal morphology–physics structure in the data. We thus have created a bidirectional, image-driven pipeline linking droplet geometry and material properties. On supervised tests, single-task model achieve $R^2 = 0.9428$ (MLP) and $0.9978$ (XGBoost) for viscosity, and $R^2 = 0.9843$ (MLP) and $0.9996$ (XGBoost). In the multi-task setting, we obtained $R^2 = 0.9361$ (MLP) and $0.9843$ (XGBoost) for viscosity, and $R^2 = 0.9854$ (MLP) and $0.9995$ (XGBoost) for surface tension. For the inverse mapping, the test-set coefficient of determination in latent space reaches $R^2 = 0.9843$ with XGBoost (MLP: 0.9145), which is significantly better than MLP model, indicating that tree-based model performs better than simple MLP in these cases. In addition, our analyses demonstrate the XGBoost models perform very well in terms of errors, with surface tension predictions achieving a maximum percentage error  $<2\%$. These results demonstrate that the instantaneous pinch-off profile encodes fluid properties with high fidelity, and that a compact latent representation can be robustly regressed from physical parameters. 
Unsupervised analyses on latent shape features identify five physically interpretable clusters. The clusters separate cleanly across the $Re$–$Oh$ and $Bo$–$Oh$ plane, indicating that the latent representation captures genuine fluid-mechanical differences. This could provide a new insight on the classification of droplet flow regimes. Our uncertainty analysis demonstrates that the models work particularly well for surface tension predictions and for high viscosities, having limited success for low viscosity liquids. 

Previous studies have demonstrated that the simultaneous measurement of surface tension and viscosity is possible using specialized techniques \cite{lohse2022fundamental}, with most traditional characterization protocols rely on separate tensiometers and viscometers, which often demand substantial sample volumes \cite{bautista2018simple} and/or are limited by their measurement range \cite{fainerman2004maximum}. Our approach provides a single-step measurement of both surface tension and viscosity, alleviating the need for multiple instruments \cite{seimiya2025simultaneous}. Our results demonstrate a generalizable route to rapid, multi-parameter characterization of liquids from a single pinch-off snapshot, which is suited for integration into automated or in-line monitoring systems. The findings open up new possibilities in inkjet printing, e.g. real-time characterization of ink viscosity and/or surface tension during jetting \cite{kim2022design, phung2023machine}, or in biomedical diagnostics with rapid small-volume property checks \cite{lenzen2024portable}, of on other industrial processes with on-the-fly materials characterization \cite{tronci2019line}. 

\textcolor{black}{We note our method is validated for gravity-driven droplet formation from circular nozzles into air and is restricted to the 0.001 $<$ Re $<$ 200 and 0.01 $<$ Oh $<$ 20 ranges. Within this parameter space the cases considered span several orders of magnitude. We agree, however, that transferability beyond this geometry is not guaranteed. Likewise, the surrounding medium is air, such that the approach is most appropriate when density and viscosity contrasts remain large and the dynamical influence of the outer fluid is relatively weak.}

By making our dataset (and code) openly available to the community, we expect that the database and models will be extended and potentially improved by including numerically obtained results and  additional experiments under conditions not attainable in our facilities, thereby broadening validation and improving generalization.

\section*{Data Availability}
The data and codes used in this study are openly available at: \url{https://github.com/CFTL-Illinois/Droplet-Properties-Prediction}.

\begin{acknowledgments}
This work was supported by UCL’s Engineering and Physical Sciences Research Council (EPSRC) Impact Acceleration Account 2022-25. AAC-P acknowledges funding support from the UK Natural Environment Research Council (UKRI1271) and the UK Engineering and Physical Sciences Research Council (EP/W016036/1). We also wish to acknowledge both reviewers and the editor, whose thoughtful comments and suggestions helped us to improve our manuscript.
\end{acknowledgments}

\appendix

\section{Convolutional Autoencoder Architecture}
\label{app:autoencoder}

The droplet break-up silhouettes used in this study contain fine-scale
interface curvatures that must be encoded in a compact numerical
representation. To achieve this, we trained a convolutional autoencoder that maps each padded grayscale droplet image of size $875\times875$ pixels into a 14-dimensional latent vector, which is later used as an input feature for the supervised and unsupervised learning models.

Table~\ref{tab:autoencoder_arch} lists the encoder–decoder parameters. The encoder progressively downsamples the image through six stride-2 convolutional layers, increasing the channel depth while reducing the spatial resolution. After flattening, a fully connected layer outputs a
latent representation $z\in\mathbb{R}^{14}$. The decoder mirrors this
structure: a fully connected expansion restores the spatial dimensions,
followed by six transposed convolution layers that reconstruct the original
$875\times875$ image.

\begin{figure}[t]
\centering
\includegraphics[width=0.9\columnwidth]{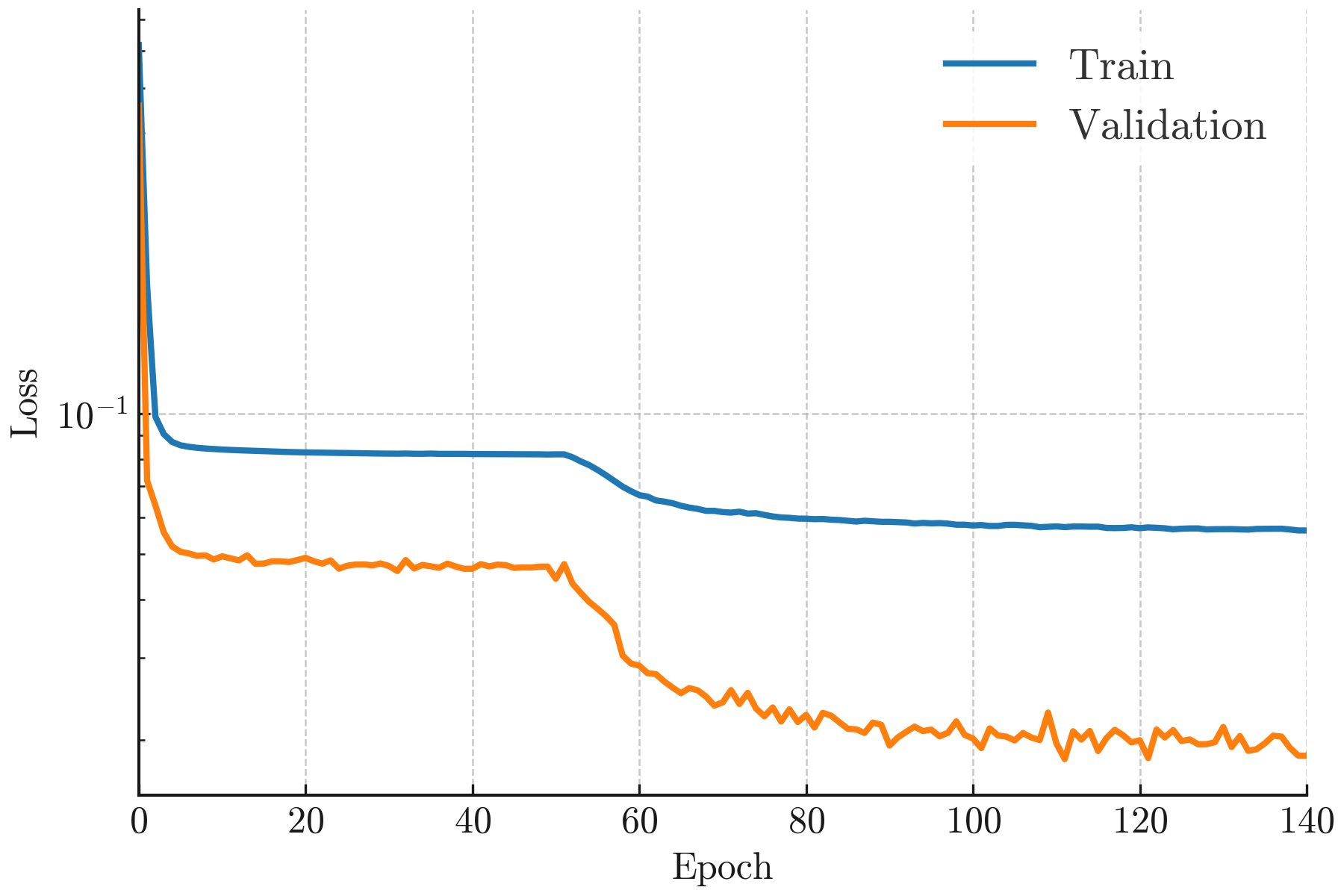}
\caption{Training and validation loss curves of the autoencoder. Early stopping is
triggered at epoch $140$.}
\label{fig:autoencoder_loss}
\end{figure}

The training and validation loss curves seen in Fig.~\ref{fig:autoencoder_loss}
indicate a stable convergence. Early stopping is triggered at epoch~140 
to obtain an accurate reconstruction without overfitting.

\begin{table*}[t]
\centering
\begin{tabular}{ll}
\toprule
\textbf{Encoder} & \textbf{Operation} \\
\midrule
Input & $875 \times 875 \times 1$ grayscale image (padded) \\
Conv Layer 1 & Conv2d(1, 32, kernel=5, stride=2, padding=2), ReLU \\
Conv Layer 2 & Conv2d(32, 64, kernel=5, stride=2, padding=2), ReLU \\
Conv Layer 3 & Conv2d(64, 128, kernel=5, stride=2, padding=2), ReLU \\
Conv Layer 4 & Conv2d(128, 256, kernel=5, stride=2, padding=2), ReLU \\
Conv Layer 5 & Conv2d(256, 256, kernel=5, stride=2, padding=2), ReLU \\
Conv Layer 6 & Conv2d(256, 256, kernel=5, stride=2, padding=2), ReLU \\
Flatten & Converts $256 \times 14 \times 14$ to vector \\
FC Layer & Linear($256 \cdot 14 \cdot 14 \rightarrow 14$) $\Rightarrow$ latent vector $z$ \\
\midrule
\textbf{Decoder} & \textbf{Operation} \\
\midrule
Input & Latent vector $z \in \mathbb{R}^{14}$ \\
FC Layer & Linear($14 \rightarrow 256 \cdot 14 \cdot 14$), ReLU \\
Reshape & $256 \times 14 \times 14$ \\
Transposed Conv 1 & ConvTranspose2d(256, 256, kernel=5, stride=2, padding=2, output\_padding=1), ReLU \\
Transposed Conv 2 & ConvTranspose2d(256, 256, kernel=5, stride=2, padding=2, output\_padding=1), ReLU \\
Transposed Conv 3 & ConvTranspose2d(256, 128, kernel=5, stride=2, padding=2, output\_padding=1), ReLU \\
Transposed Conv 4 & ConvTranspose2d(128, 64, kernel=5, stride=2, padding=2, output\_padding=1), ReLU \\
Transposed Conv 5 & ConvTranspose2d(64, 32, kernel=5, stride=2, padding=2, output\_padding=1), ReLU \\
Transposed Conv 6 & ConvTranspose2d(32, 1, kernel=6, stride=2, padding=2, output\_padding=1), Sigmoid \\
Output & Reconstructed image ($875 \times 875 \times 1$) \\
\bottomrule
\end{tabular}
\caption{Convolutional autoencoder architecture used for the droplet silhouette compression. The encoder reduces the input image to a 14-dimensional latent vector, and the decoder reconstructs the full-resolution image from the latent vector.}
\label{tab:autoencoder_arch}
\end{table*}


\section{Supervised Learning Models}
\label{app:supervised}
\begin{figure*}[t]
    \centering
    \vspace{-5mm}
    \includegraphics[width=0.9\textwidth]{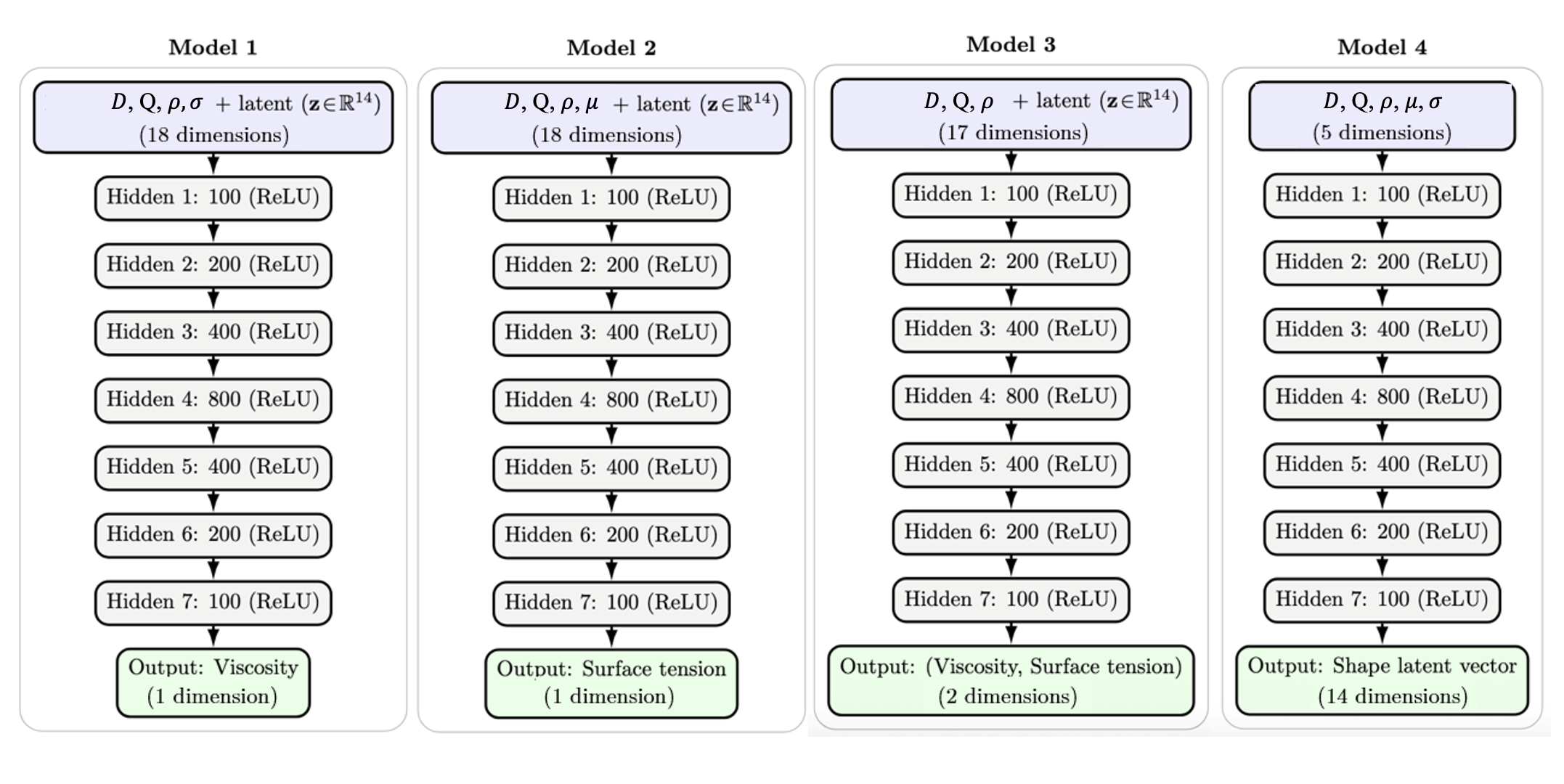}
    \caption{Architecture summary of the supervised MLP models used in this work.
Depending on the task, the inputs consist of physical parameters and, when required, the 14-dimensional latent droplet representation. The models predict viscosity (model 1), surface tension (model 2), viscosity and surface tension (model 3), or the 14-dimensional shape descriptor (model 4).}
    \label{fig:supervised_diagram}
\end{figure*}

\begin{figure*}[t]
    \centering
    \includegraphics[width=0.9\textwidth]{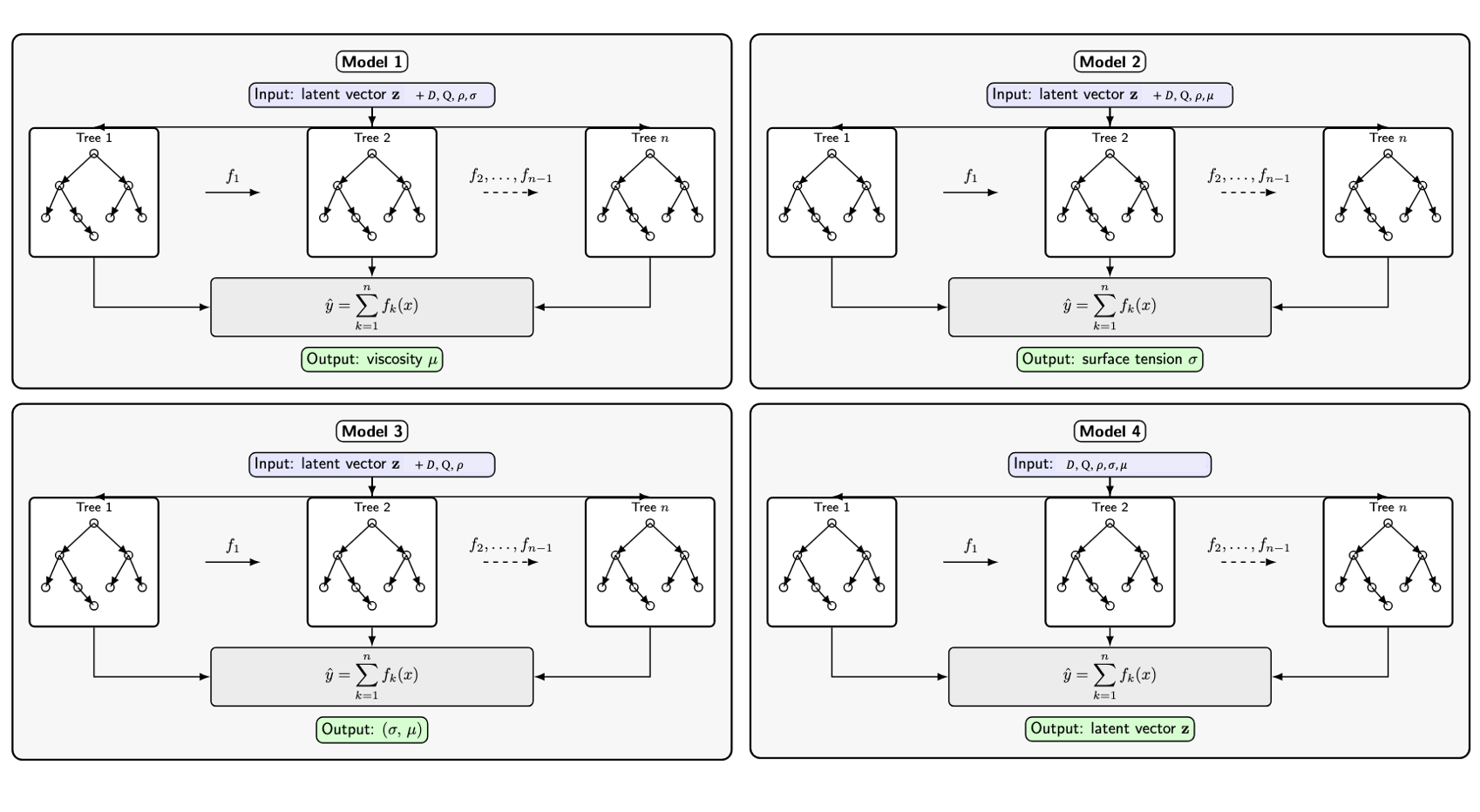}
    \caption{Supervised learning framework (XGBoost). Model inputs consist of physical parameters and, when required, the 14-dimensional latent droplet representation. The models predict viscosity (Model 1), surface tension (Model 2), viscosity and surface tension (Model 3), or the 14-dimensional shape descriptor (Model 4).}
    \label{fig:supervised_diagram_xgb}
\end{figure*}

Four supervised models were trained to map droplet images and physical parameters to quantitative
fluid properties. Models 1–3 predict viscosity and/or surface tension from measured physical variables and the 14-dimensional latent vectors produced by the convolutional autoencoder. In contrast, Model 4 performs the inverse task: it predicts the droplet’s latent geometric descriptor directly from five physical variables, enabling comparisons between
experimental outcomes and shape embeddings. Fig.~\ref{fig:supervised_diagram} shows a schematic multilayer perceptron (MLP) framework used for Models 1–4.  Each model uses a fully connected architecture with seven hidden layers and ReLU activation functions, but differs in the dimensionality of its input and output spaces. \textcolor{black}{Meanwhile, Fig.~\ref{fig:supervised_diagram_xgb} shows a XGBoost framework used for Models 1–4.}

\section{Clustering Algorithms}
\label{app:clustering_algorithms}
Here, we summarize the mathematical formulation of the clustering methods, i.e. the $K$-means and the Gaussian mixture (GMM) algorithms.  
Both methods operate on the latent representations $z_i \in \mathbb{R}^{14}$ that were
extracted from the convolutional autoencoder.

\subsection{$K$-Means Clustering}

The $K$-means algorithm partitions latent vectors $\{z_i\}_{i=1}^N$ into $K$
clusters by minimizing the within-cluster sum of squared distances:
\begin{equation}
J(\{\mu_k\}, \{c_i\})
=
\sum_{i=1}^{N}
\| z_i - \mu_{c_i} \|^2,
\end{equation}
where  
$c_i \in \{1,\ldots,K\}$ is the assignment of sample $i$  
and  
$\mu_k \in \mathbb{R}^d$ is the centroid of cluster $k$.

\subsubsection{Assignment step}

Each data point is assigned to the nearest centroid:
\begin{equation}
c_i^{(t+1)} =
\arg\min_k
\| z_i - \mu_k^{(t)} \|^2.
\end{equation}

\subsubsection{Update step}

Each centroid is updated as the mean of its assigned samples:
\begin{equation}
\mu_k^{(t+1)} =
\frac{
\sum_{i:\,c_i^{(t+1)} = k} z_i
}{
|\{i : c_i^{(t+1)} = k\}|
}.
\end{equation}

Iterations terminate when neither assignments nor centroids change appreciably.

\subsection{Gaussian Mixture Models and the EM Algorithm}

A Gaussian mixture model with $K$ components represents the probability density
of a latent vector $z \in \mathbb{R}^d$ as
\begin{equation}
p(z \mid \Theta)
=
\sum_{k=1}^{K}
\pi_k \,
\mathcal{N}(z \mid \mu_k, \Sigma_k),
\end{equation}
where $\pi_k$ are the mixture weights ($\pi_k \ge 0$, $\sum_k \pi_k = 1$),  
$\mu_k \in \mathbb{R}^d$ are the component means, and $\Sigma_k \in \mathbb{R}^{d \times d}$ are covariance matrices. The parameters $\Theta = \{\pi_k, \mu_k, \Sigma_k\}_{k=1}^{K}$ are estimated by
maximizing the log-likelihood
\begin{equation}
\mathcal{L}(\Theta)
=
\sum_{i=1}^{N}
\log p(z_i \mid \Theta).
\end{equation}

Because the likelihood admits no closed-form maximizer, the EM algorithm is used to construct a sequence of estimates $\Theta^{(t)}$ that monotonically increases $\mathcal{L}$.

\subsubsection{E-step}

The posterior responsibility assigned to each mixture component $k$ for a data point $z_i$ is given by
\begin{equation}
\gamma_{ik}^{(t)} =
\frac{
\pi_k^{(t)} 
\, \mathcal{N}(z_i \mid \mu_k^{(t)}, \Sigma_k^{(t)})
}{
\sum_{j=1}^{K}
\pi_j^{(t)} 
\, \mathcal{N}(z_i \mid \mu_j^{(t)}, \Sigma_j^{(t)})
},
\end{equation}
with these values satisfying $0 \le \gamma_{ik} \le 1$ and
$\sum_{k=1}^{K} \gamma_{ik} = 1$.

\subsubsection{M-step}

The effective membership of component $k$ is
\begin{equation}
N_k^{(t+1)} = \sum_{i=1}^N \gamma_{ik}^{(t)}.
\end{equation}

The updated mixture weights, means, and covariances are given by: 
\begin{align}
\pi_k^{(t+1)} &= \frac{N_k^{(t+1)}}{N}, \\
\mu_k^{(t+1)} &=
\frac{
\sum_{i=1}^{N} \gamma_{ik}^{(t)} z_i
}{N_k^{(t+1)}}, \\
\Sigma_k^{(t+1)} &=
\frac{
\sum_{i=1}^{N}
\gamma_{ik}^{(t)}
(z_i - \mu_k^{(t+1)})
(z_i - \mu_k^{(t+1)})^T
}{N_k^{(t+1)}}.
\end{align}

Iterations continue until the increase in log-likelihood drops below a specified threshold.
\section{Cluster Evaluation Results}
\label{app:cluster_evaluation}

We evaluated two standard metrics for both the K-Means
and the Gaussian mixture models. The first is Silhouette Score \cite{rousseeuw1987silhouettes}
\begin{equation*}
s(z_i) = \frac{b(z_i) - a(z_i)}{\max\{a(z_i),\, b(z_i)\}},
\end{equation*}
where $a(z_i)$ denotes the mean distance from $z_i$ to other points in the same cluster, and $b(z_i)$ is the mean distance to points in the nearest different cluster.

The second is Davies–Bouldin Index \cite{davies2009cluster}
\begin{equation*}
DBI = \frac{1}{K} \sum_{i=1}^{K} \max_{j \neq i} \left( \frac{s_i + s_j}{d_{ij}} \right),
\end{equation*}
where $s_i$ is the average distance between points in cluster $i$ and its centroid, and $d_{ij}$ is the distance between the centroids of clusters $i$ and $j$.These metrics provide complementary assessments of how well the data partition into distinct clusters for a given choice of K. 

A higher Silhouette Score, approaching 1, indicates that the clusters are well-separated and compact. Lower DBI values indicate better clustering performance, favoring compact and well-separated clusters.

\begin{figure}[t]
\centering

\begin{tikzpicture}
    \node[anchor=south west, inner sep=0] at (0,0)
        {\includegraphics[width=0.8\linewidth]{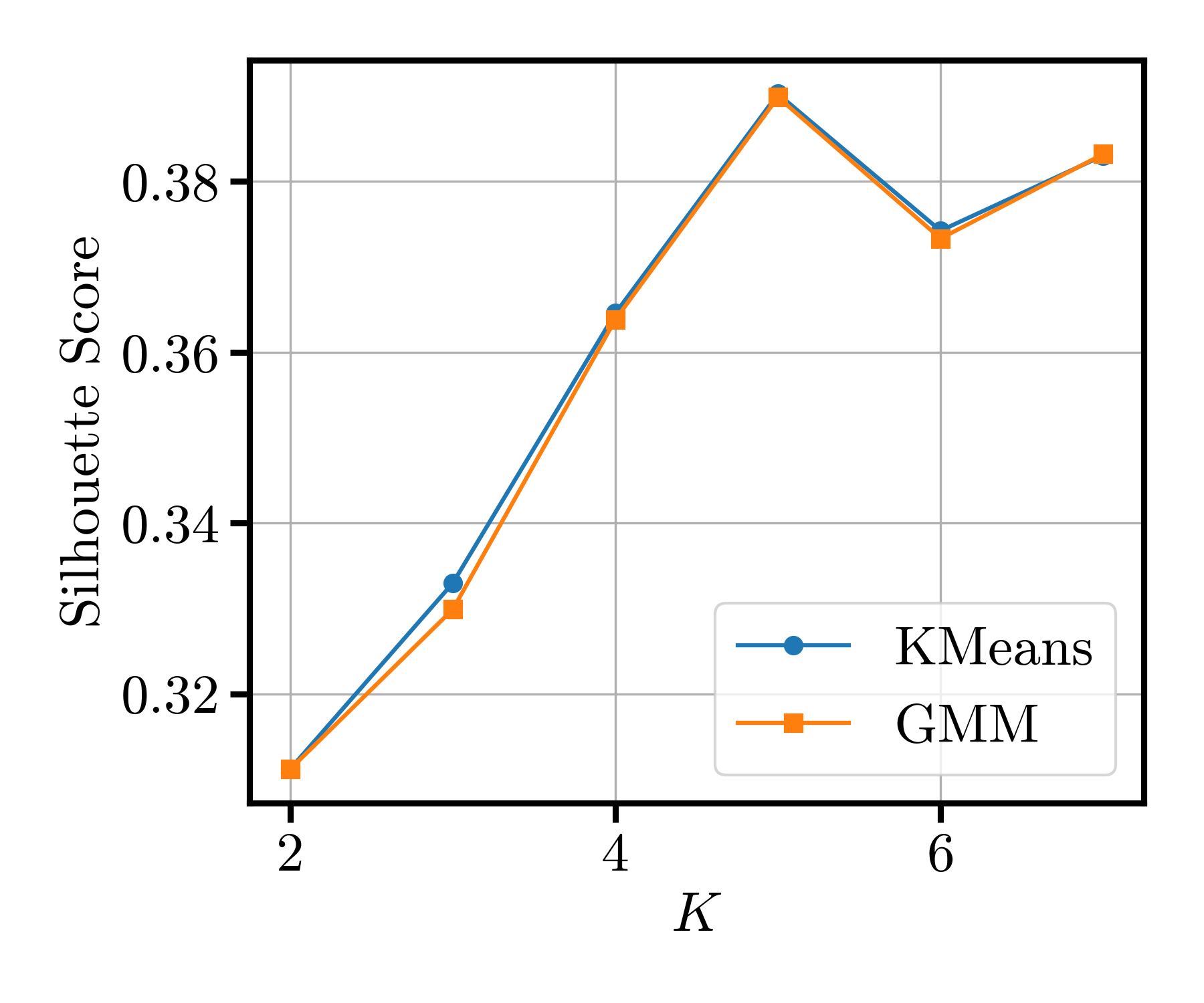}};
    \node[anchor=south west, inner sep=2pt] at (0,0.5) {\scriptsize \textbf{(a)}};
\end{tikzpicture}

\vspace{-5mm}

\begin{tikzpicture}
    \node[anchor=south west, inner sep=0] at (0,0)
        {\includegraphics[width=0.8\linewidth]{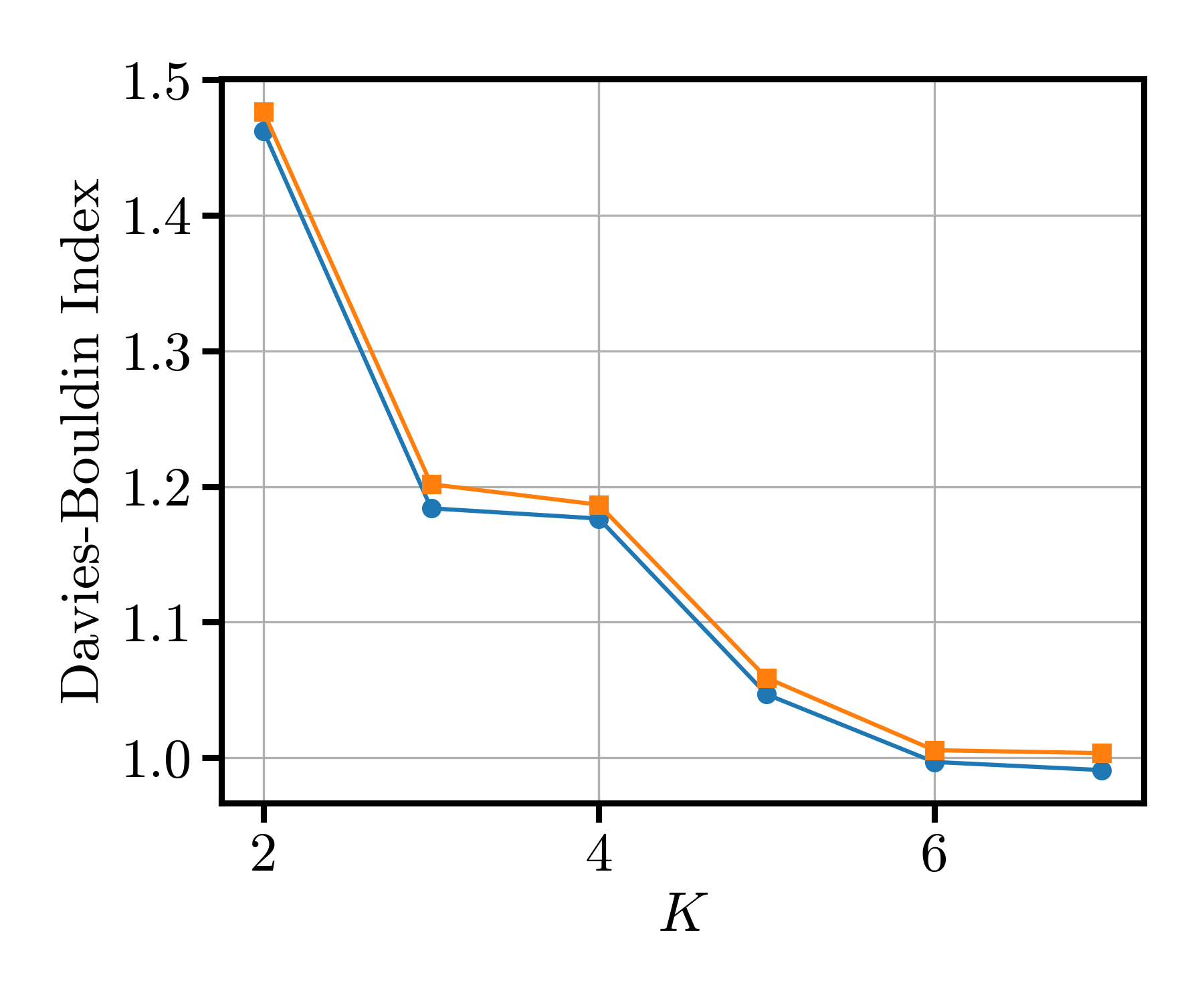}};
    \node[anchor=south west, inner sep=2pt] at (0,0.5) {\scriptsize \textbf{(b)}};
\end{tikzpicture}

\caption{
Cluster evaluation results.  
\textbf{(a)} Silhouette Scores for K-Means and GMM across cluster sizes $K$.  
\textbf{(b)} Davies–Bouldin Index (DBI) for K-Means and GMM across $K$.  
}
\label{fig:silhouette_dbi_hdbscan}
\end{figure}

Figure~\ref{fig:silhouette_dbi_hdbscan}a shows the Silhouette Score as a function of the cluster  number $K$. The score measures the relative separation between
clusters: values closer to~1 indicate samples that are well matched to their own
cluster and poorly matched to neighboring clusters. Both K-Means and GMM exhibit nearly identical trends, with the score increasing from $K=2$ to $5$ and reaching a broad maximum in this range. This behavior suggests that the learned latent manifold naturally organizes the  droplet silhouettes into a moderate number of distinct morphological families. Increasing $K$ beyond this range yields only marginal improvement, indicating that additional splitting does not reveal further meaningful structure.

Figure~\ref{fig:silhouette_dbi_hdbscan}b presents the DBI, which quantifies the average similarity between each cluster and its most similar neighbor.   Lower DBI values correspond to better-defined, more separated clusters. Consistent with the Silhouette analysis, the DBI decreases sharply from $K=2$ to $K=5$–6 for both algorithms, after which improvements saturate. The close agreement between K-Means and GMM across both metrics reinforces the stability of the cluster structure and indicates that the latent space provides a robust embedding in which fluid composition and interfacial behavior are reflected as geometric regularities. 

These two metrics show that $K \approx 5$ provides an optimal balance between cluster compactness and separation, matching the visual grouping observed in the  latent-space projections and supporting the use of five-cluster partitioning in the analysis of droplet break-up regimes.

\section{K-means and GMM Cluster Assignments}
\label{app:cluster}

This section reports the full cluster assignment counts for the liquids tested in the dataset. 
Tables~\ref{tab:kmeans_liquid_cluster} and \ref{tab:gmm_liquid_cluster} provide a fine-grained view of liquid formulation distributions across the five clusters obtained from the K-means and Gaussian mixture models. As observed, highly viscous liquids---including pure glycerin and silicone oils of
350\,cSt and 1000\,cSt--- largely occupy cluster 4, reflecting their
well-separated break-up behavior in the latent and physical parameter spaces.
Conversely, low-viscosity mixtures largely populate cluster 0, consistent with their overlapping regimes in the $Bo$–$Oh$ and $Re$–$Oh$ diagrams.

\begin{table}[t]
\centering

\begin{tabular}{lrrrrr}
\toprule
 & \multicolumn{5}{c}{\textbf{cluster}} \\

\textbf{Liquid type} & 0 &  1 &  2 &  3 &  4 \\
\midrule
10\% water + 90\% ethanol & 10 & 5 & 5 & 10 & 0 \\
10\% water + 90\% glycerin & 1 & 5 & 0 & 5 & 14 \\
20\% water + 80\% glycerin & 18 & 8 & 18 & 10 & 0 \\
30\% water + 70\% ethanol & 10 & 5 & 5 & 10 & 0 \\
30\% water + 70\% glycerin & 5 & 5 & 10 & 10 & 0 \\
40\% water + 60\% glycerin & 8 & 9 & 18 & 19 & 0 \\
5\% water + 95\% ethanol & 10 & 5 & 5 & 10 & 0 \\
5\% water + 95\% glycerin & 4 & 5 & 0 & 1 & 15 \\
50\% water + 50\% ethanol & 10 & 5 & 5 & 10 & 0 \\
50\% water + 50\% glycerin & 0 & 10 & 13 & 18 & 0 \\
70\% water + 30\% ethanol & 10 & 5 & 9 & 6 & 0 \\
70\% water + 30\% glycerin & 0 & 10 & 12 & 10 & 0 \\
80\% water + 20\% ethanol & 14 & 9 & 18 & 18 & 0 \\
90\% water + 10\% ethanol & 4 & 5 & 10 & 11 & 0 \\
90\% water + 10\% glycerin & 0 & 10 & 11 & 10 & 0 \\
Pure glycerin & 0 & 0 & 0 & 2 & 28 \\
Pure methanol & 19 & 9 & 9 & 17 & 0 \\
Pure ethanol & 20 & 9 & 9 & 16 & 0 \\
Pure water & 0 & 16 & 10 & 10 & 0 \\
Silicone oil 1000 cSt & 0 & 0 & 0 & 0 & 8 \\
Silicone oil 2 cSt & 15 & 5 & 5 & 5 & 0 \\
Silicone oil 350 cSt & 0 & 0 & 0 & 0 & 18 \\
Silicone oil 50 cSt & 0 & 9 & 9 & 0 & 36 \\
Silicone oil 5 cSt & 17 & 5 & 5 & 3 & 0 \\
\bottomrule
\end{tabular}

\caption{K-means cluster assignment counts for all liquid types.}
\label{tab:kmeans_liquid_cluster}
\end{table}

\begin{table}[t]
\centering

\begin{tabular}{lrrrrr}
\toprule
 & \multicolumn{5}{c}{\textbf{cluster}} \\
\textbf{Liquid type} & 0 &  1 &  2 &  3 &  4 \\
\midrule
10\% water + 90\% ethanol & 10 & 5 & 5 & 10 & 0 \\
10\% water + 90\% glycerin & 1 & 5 & 0 & 5 & 14 \\
20\% water + 80\% glycerin & 18 & 9 & 18 & 9 & 0 \\
30\% water + 70\% ethanol & 10 & 5 & 5 & 10 & 0 \\
30\% water + 70\% glycerin & 5 & 5 & 10 & 10 & 0 \\
40\% water + 60\% glycerin & 8 & 9 & 18 & 19 & 0 \\
5\% water + 95\% ethanol & 10 & 5 & 5 & 10 & 0 \\
5\% water + 95\% glycerin & 4 & 5 & 0 & 1 & 15 \\
50\% water + 50\% ethanol & 10 & 5 & 5 & 10 & 0 \\
50\% water + 50\% glycerin & 0 & 10 & 13 & 18 & 0 \\
70\% water + 30\% ethanol & 10 & 5 & 9 & 6 & 0 \\
70\% water + 30\% glycerin & 0 & 10 & 12 & 10 & 0 \\
80\% water + 20\% ethanol & 14 & 9 & 18 & 18 & 0 \\
90\% water + 10\% ethanol & 4 & 5 & 10 & 11 & 0 \\
90\% water + 10\% glycerin & 0 & 10 & 11 & 10 & 0 \\
Pure glycerin & 0 & 0 & 0 & 2 & 28 \\
Pure methanol & 18 & 9 & 9 & 18 & 0 \\
Pure ethanol & 18 & 9 & 9 & 18 & 0 \\
Pure water & 0 & 16 & 10 & 10 & 0 \\
Silicone oil 1000 cSt & 0 & 0 & 0 & 0 & 8 \\
Silicone oil 2 cSt & 15 & 5 & 5 & 5 & 0 \\
Silicone oil 350 cSt & 0 & 0 & 0 & 0 & 18 \\
Silicone oil 50 cSt & 0 & 9 & 9 & 0 & 36 \\
Silicone oil 5 cSt & 17 & 5 & 5 & 3 & 0 \\
\bottomrule
\end{tabular}
\caption{GMM cluster assignment counts for all liquid types.}
\label{tab:gmm_liquid_cluster}
\end{table}

\bigskip

\section{Error Analysis}
\label{app:error}

\begin{figure}[t]
\centering

    \begin{tikzpicture}
        \node[anchor=south west, inner sep=0] at (0,0)
            {\includegraphics[width=\linewidth]{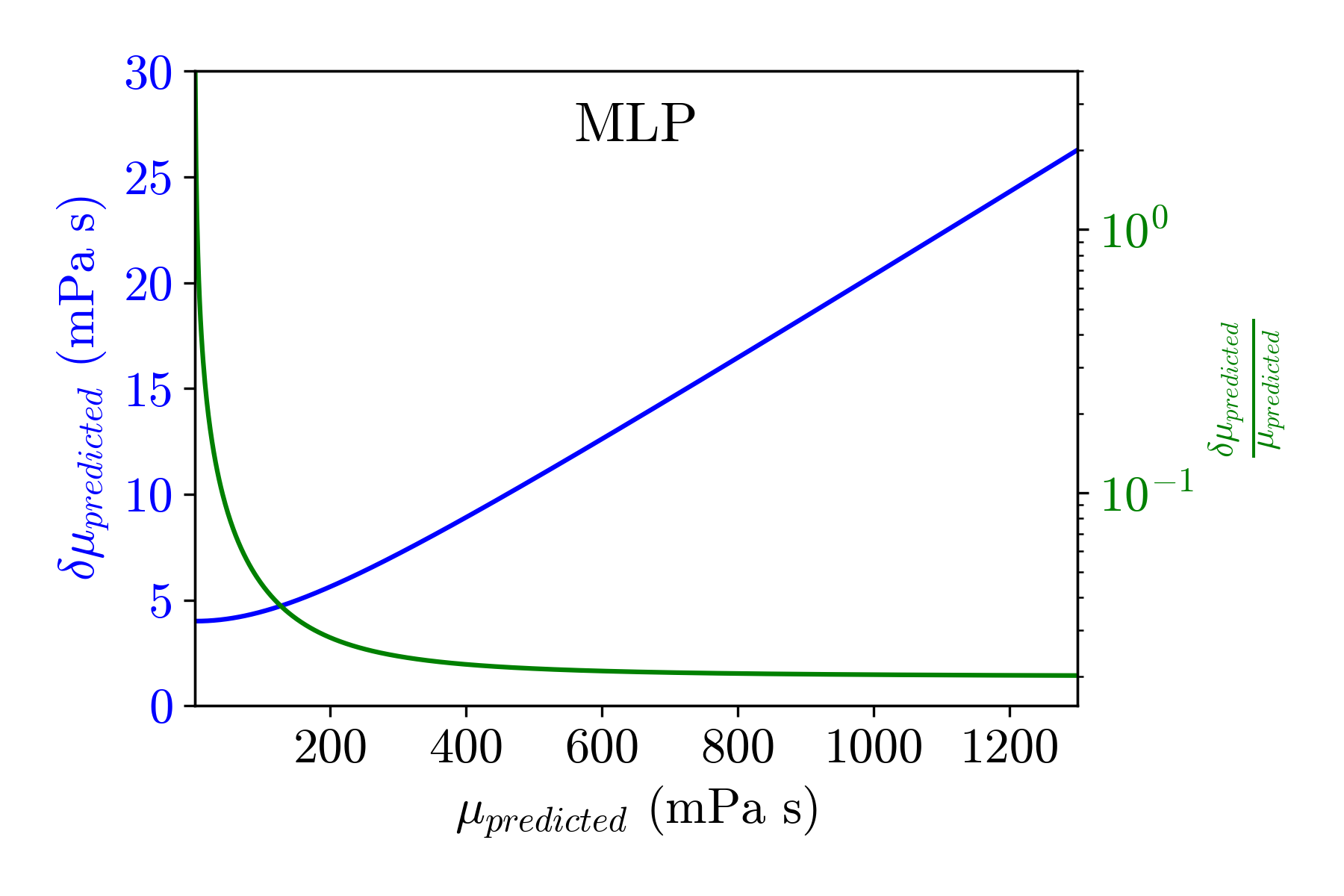}};
        \node[anchor=south west, inner sep=2pt] at (0,0.5) {\scriptsize \textbf{(a)}};
    \end{tikzpicture}

\vspace{-3mm}

\begin{tikzpicture}
    \node[anchor=south west, inner sep=0] at (0,0)
        {\includegraphics[width=\linewidth]{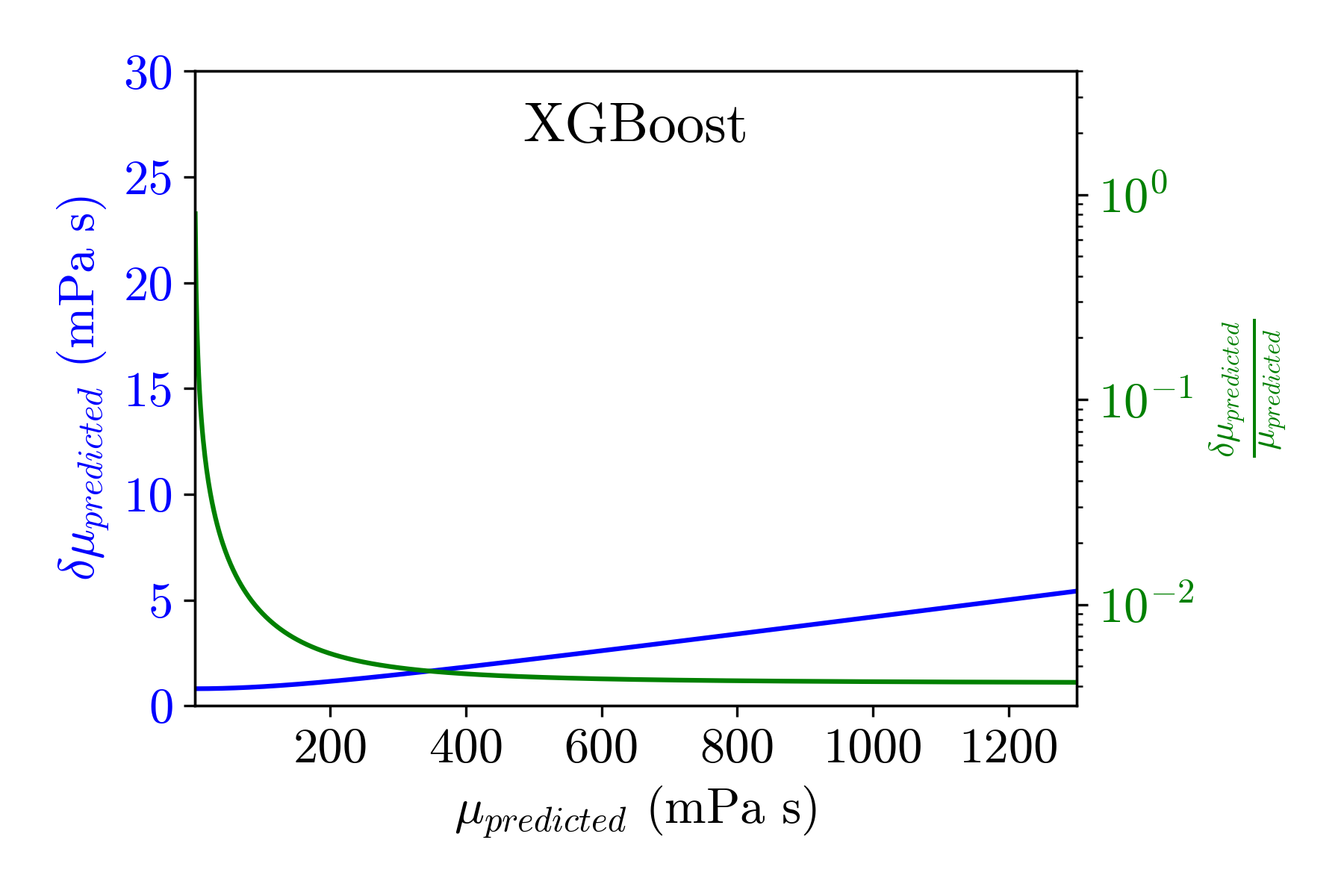}};
    \node[anchor=south west, inner sep=2pt] at (0,0.5) {\scriptsize \textbf{(b)}};
\end{tikzpicture}

\begin{tikzpicture}
    \node[anchor=south west, inner sep=0] at (0,0)
        {\includegraphics[width=\linewidth]{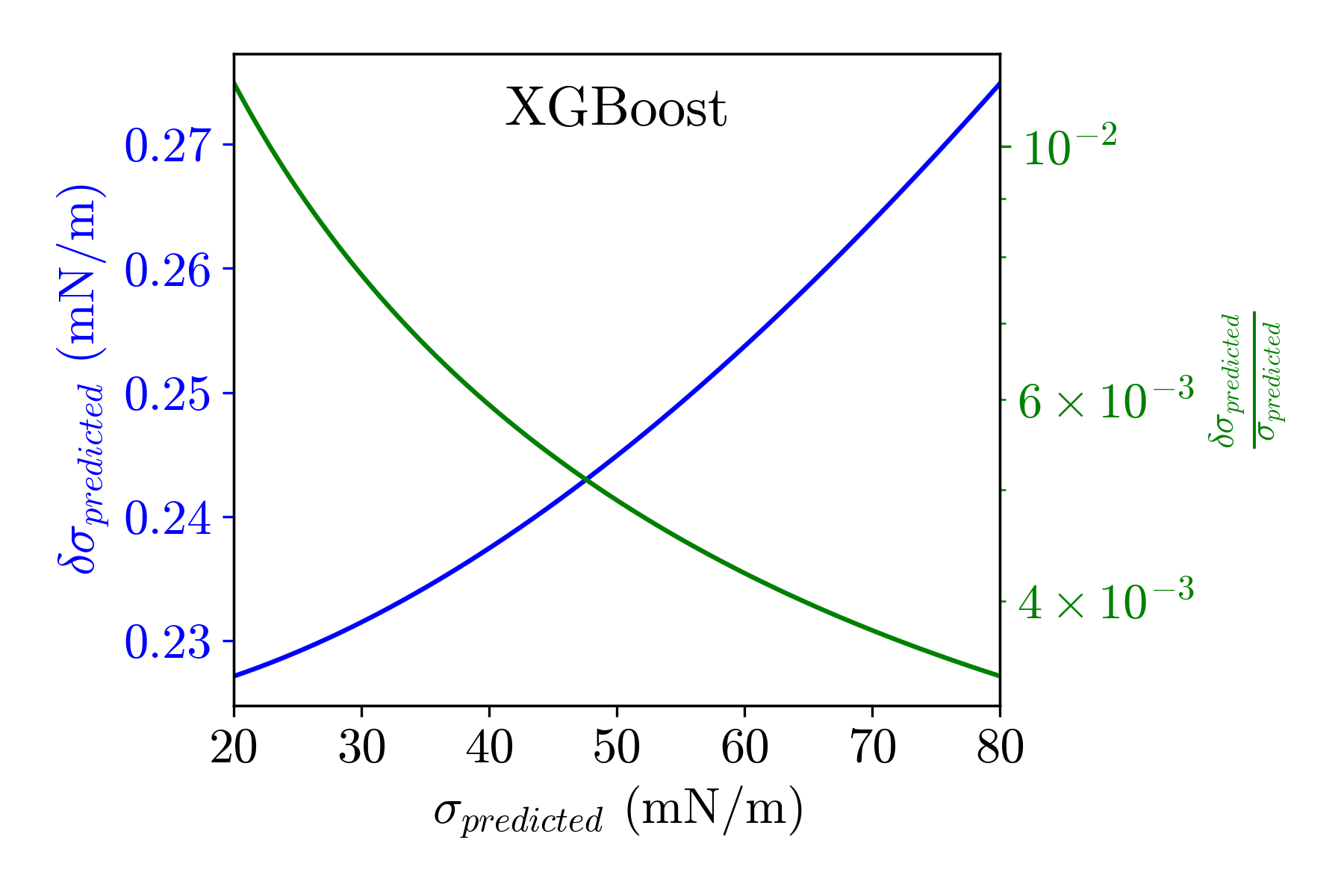}};
    \node[anchor=south west, inner sep=2pt] at (0,0.5) {\scriptsize \textbf{(c)}};
\end{tikzpicture}

\caption{Overall error and fractional error in terms of the predicted value.   
\textbf{(a)} Prediction error in terms of the viscosity value; MLP method. 
\textbf{(b)} Prediction error in terms of the viscosity value; XGBoost method.
\textbf{(c)} Prediction error in terms of the surface tension value; XGBoost method.
}
\label{fig:error}
\end{figure}

\clearpage
\bibliography{Main}

\end{document}